\documentclass[graybox]{svmult}

\usepackage{mathptmx}       % selects Times Roman as basic font
\usepackage{helvet}         % selects Helvetica as sans-serif font
\usepackage{courier}        % selects Courier as typewriter font
\usepackage{type1cm}        % activate if the above 3 fonts are
\usepackage{makeidx}         % allows index generation
\usepackage{graphicx}        % standard LaTeX graphics tool
\usepackage{multicol}        % used for the two-column index
\usepackage[bottom]{footmisc}% places footnotes at page bottom
\usepackage{amsfonts}
\usepackage{color}
\usepackage{cite}
\makeindex             % used for the subject index
\begin{document}
%\SetWatermarkText{Tomate}
%\SetWatermarkText{\includegraphics[width=\textwidth]{figure.file}}
%\SetWatermarkScale{1.5} %default 1.2 (5 cm)
%\SetWatermarkLightness{0.92} %0=black; 1_grey
\title{Complex networks analysis in socioeconomic models}
\author{Luis M. Varela, Giulia Rotundo, Marcel Ausloos, Jes\'us Carrete}
\institute{Luis M. Varela \at Grupo de Nanomateriais e Materia Branda.\\ Departamento de F\'{\i}sica da Materia Condensada, Univ. de Santiago de Compostela, Campus Vida s/n. E-15782 Santiago de Compostela. Spain, \email{luismiguel.varela@usc.es}
\and Giulia Rotundo \at Department of Methods and Models for Economics, Territory and Finance. La Sapienza University of Rome, via del Castro Laurenziano 9, I-00161 Rome, Italy \email{giulia.rotundo@gmail.com}
\and Marcel Ausloos \at R\' es. Beauvallon, r. Belle Jardini\`ere, 483/0021, B-4031 Li\`ege, Wallonia-Brussels Federation \email{marcel.ausloos@ulg.ac.be}
\and Jes\'us Carrete \at  Grupo de Nanomateriais e Materia Branda.\\ Departamento de F\'{\i}sica da Materia Condensada, Univ. de Santiago de Compostela, Campus Vida s/n. E-15782 Santiago de Compostela. Spain. \\ \& LITEN, CEA-Grenoble, 17 rue des Martyrs, BP166, F-38054, Grenoble, Cedex 9, France \email{jcarrete@gmail.com}}
%\marginpar{Please Marcel, reconsider the address for final publication}
%\author{Luis M. Varela \inst{1},  Jes\'us Carrete \inst{1}$^{,}$\inst{2}, Giulia Rotundo\inst{3}\\ \and Marcel Ausloos\inst{4}}

%\institute{
%\and

\maketitle

\begin{center}
\today
\end{center}
\abstract{This chapter aims at reviewing complex networks models and methods that were either developed for or applied to socioeconomic issues, and pertinent to the theme of New Economic Geography.
After an introduction to the foundations of the field of complex networks, the present summary adds insights on the statistical mechanical approach, and on the most relevant computational aspects for the treatment of these systems. As the most frequently used model for interacting agent-based systems, a brief treatment of the statistical mechanics of the classical Ising model on regular lattices, together with recent treatments of the same model on small-world Watts-Strogatz and scale-free Albert-Barab\'asi complex networks is included.
Other sections of the chapter are devoted to applications of complex networks to economics, finance, spreading of innovations, and regional trade and developments. The chapter also reviews results involving applications of complex networks to other relevant socioeconomic issues, including results for opinion and citation networks. Finally, some avenues for future research are introduced before summarizing the main conclusions of the chapter.}

%\end{abstract}

%
\section{Introduction}

The foundation of the field of network topology dates back to the 18th century with the seminal work in graph theory of Euler \cite{euler} devoted to the celebrated problem of  K\"onigsberg bridges, and includes several important contributions in the last two centuries like Cayley trees \cite{Cayley} and the theory of random graphs due to Erd\"os and R\' enyi \cite{Erdos-Renyi}. However, it was not until the late nineties that complex networks with specific structural features valid for the description of short path lengths, highly clustered \cite{SWN}, and even heterogeneous networks \cite{AB1999} were introduced, which opened what could be called the contemporary era of network theory. The amount of papers published since then has never ceased to increase exponentially as network theory started to be applied to fields like physics, biology, computer science, sociology, epidemiology, and economics among others. It is now recognized that a network is always the skeleton of any complex system, so it is by no means an exaggeration to say that network theory has become one of the cornerstones of the theory of complex systems. All this activity has been excellently reviewed in several occasions during the last decade (see for e.g. Refs. \cite{reviewAB,newman0303516,PastorSatorrasRDG,reviewBocaletti,reviewnewmanwatts}), and it is now widely accepted that the dynamics of many complex systems corresponds to emergent phenomena associated to the large scale fluctuations of some real network. 

The effects of the topological properties of networks on dynamical processes is also a matter of intense research inside the field of complex networks. Some of these dynamical processes are the evolution of the network itself, spreading processes in agent-based systems (epidemics in a population, rumor spreading), opinion formation, cultural assimilation,  voting processes, or decision making on competing for limited resources (for a review, see Boccaletti et al. \cite{reviewBocaletti}). Specifically, the description of the evolution of the network is very important on itself, as real networks evolve in time. For that, one has to follow the evolution of networks, through the number of vertices, the number, weight and direction of links, and through other characteristic quantities mentioned below, as seen in \cite{reviewBocaletti,496RLMAwcss,520MARLdrastic,511RLMAj}.

In the field of economics and economic geography, a complete understanding of economic dynamics requires the understanding of its agent-based underlying structure and the interactions that give rise to the observed emergent spatial and temporal organizations, which are definitively more than the sum of its individual components (for further details see e.g. Ausloos and coworkers in this book \cite{Ugomerloneetalchapter}) %(\colorbox{yellow}{Reference to chapter by Merlone et al.}) 
The view of the economy as an evolving complex agent-based system, hereafter identified with a network, is currently gaining consensus among the scientific community (see e.g. Refs. \cite{NamatameKA,AndersonArrowPines,BlumeDurlauf}). On the basis of this network-based structural view, the application of the methods of physics of complex systems (statistical physics, nonlinear physics, and so forth) has allowed to gain new insight on the economic realm, even leading to the foundation of a new branch of Physics itself, the so called Econophysics. 
 
The purpose of this chapter is to summarize some applications that complex network theory has found in economic and social studies. Before reviewing some applications of complex networks to economic issues (financial markets, spreading of innovations, economic geography, regional trade and development....), the main results that have been developed up to now in the field of statistical mechanics of complex networks and their computational analysis will be briefly summarized. The chapter is closed with complex-network-based descriptions of other social networks, a brief description of future trends in the field of socio-economic applications of complex networks and our conclusions.

\section{Summary of statistical mechanics of complex networks}

As mentioned previously, in an attempt to reproduce the success in describing regular systems (solid state physics, phase transitions and so forth), statistical mechanics has been applied to heterogeneous systems through the formalism of complex networks, representing the constituents by means of vertices and their interactions by a set of edges. The description of these objects involves their topology and dynamic evolution, as well as different dynamic processes that take place over them. One consideration consists in tying the structure of the network and its intrinsic dynamics \cite{newman0303516}. Another concerns  the structural changes in the network due to the dynamical processes themselves. As we have previously mentioned, several excellent reviews have been published during the past decade on the structure and dynamics of complex networks, as well as on dynamic processes that take place in these topological objects. However, for the interest of the reader, in this section the main topics of the field will be briefly summarized.

%\colorbox{yellow}{Giulia: could you please include something about connectivity matrix?}
From a formal point of view a network (graph) is a pair $(V,E)$, where $V$ is a set of nodes (vertices), and $E$ is a set of links (edges), which are identified by two nodes that represent the source and the end of the link (edge).  The most used method for the representation of networks with $N$ nodes is the \emph{adjacency matrix} $A=(a_{ij})\in \mathbb{R}^{N \times N}$, whose rows and columns represent the nodes of the network, and whose terms $a_{ij}> 0$ represent the weight of the link from node $i$ to node $j$. The absence of links is given by zero elements $a_{ij}=0$. Therefore, properties of the network are reflected in the properties of the adjacency matrix. Network links are called directed if matrix A is not symmetric, which means that there exists at least one pair of indices $i, j$ such that $a_{ij} \neq a_{ji}$. This includes as particular case graphs with  upper triangular adjacency matrices, i.e. such that $a_{ij}> 0$ and $a_{ji}=0$. In graph theory, such networks are named \emph{directed acyclic graphs}. Undirected networks are represented by symmetric matrices ($A=A^T$).
A source $i$ is a node having only outgoing links (i.e. $\forall k \in V$ there are no pairs $(k,i)\in E$). By contrast, a sink $j$ is a node having only incoming links (i.e. $\forall k \in V$ there are no pairs $(j,k)\in E$).
A path from node $i_{h_1}$ to node $i_{h_k}$ in the network is a sequence of nodes $i_{h_1}, i_{h_2}, \cdots, i_{h_k}$ such that $a_{i_{h_j}, i_{h_j+1}}\neq 0$, and can be detected through the power of the adjacency matrix $A^{h_k}$.

Paths are relevant for studying diffusion processes as well as the relevance of nodes.
A network is called \emph{(strongly) connected} if a path exists between any pair of nodes, considering the direction of the links, and it is termed \emph{(weakly) connected} if a path exists among any pair of nodes when ignoring the direction of the links, which requires that the adjacency matrix be symmetric. In some applications the weight assigned to each link is 1, since the presence of the edge is relevant, but not its specific weight. 
In this case the matrix $A$ is symmetrized setting $a_{ji}=1$ for all nodes $i$, $j$ such that $a_{ij}=1$.
In applications of networks to problems in which the weight of the link is relevant, whether symmetrization is applied depends on the objectives. 

The main difference between network theory and graph theory lies in their targets. Naturally, algorithms first developed in the field of graph theory are useful and currently used also for complex networks and in other fields related to dynamical systems and the discretization of maps, like symbolic images \cite{Avrutin,symbolic}.

\subsection{Main measures for complex networks}

Networks can be classified according to a relatively large number of criteria. Taking into account the distribution of the number of links per node, networks can be classified as purely random networks (Poisson distributed), exponentially distributed small-world networks, and scale-free networks. According to the directionality of contacts they can be classified directed or undirected graphs. According to the heterogeneity in the capacity and the intensity of the connections, networks can be classified into weighted or unweighted networks depending on whether different weights are associated to their edges. Sparse and fully connected networks differ in the fraction of interconnected nodes. Finally, depending on their time evolution, networks can be classified into static and evolving.

The description of the topology of complex networks is based on several concepts and parameters that measure different features of these topological structures and macroscopic characteristics of the networks (a more detailed treatment can be found in \cite{daCostareview}). The most important of these are:

%\bibitem{daCostareview} L. DA F. COSTA*, F. A. RODRIGUES, G. TRAVIESO and P. R. VILLAS BOAS, Characterization of complex networks: A survey of measurements. Advances in Physics, Vol. 56, No. 1, February 2007, 167–242

\begin{enumerate}
\item{Average path length: no proper metric space can be defined for complex networks (usually hidden metric spaces are defined \cite{hidden}), and the (chemical) distance between any two vertices $l_{ij}$ is defined to be the number of steps from one point to the other following the shortest path. In many real networks, the average distance between two nodes, i.e. the average path-length, $<l>$, is relatively small as compared to the total number of nodes in the network. In fact, in a regular lattice,  a topological structure which can be generated from a basis for the vector space by forming all linear combinations with integer coefficient and where all the nodes are connected to the same number of neighbors, the average path length scales with the number of nodes in the network, $N$, as $<l>\sim\sqrt{N}$, while in a small-world network with long-distance shortcuts, $<l>\sim \log{N}$, so the separation between any two nodes is usually very small. This property is behind the small-world concept first studied by Milgram in his 1967 seminal paper \cite{Milgram1967}, and it is by no means included in conventional regular lattices.  The connectivity can also be measured by means of the diameter of the graph, $d$, defined as the maximum distance between any pair of its nodes. Djikstra algorithm \cite{Djikstra} is the most often used for this calculation.}

\item{Degree distribution, $p(k)$, measuring the probability that a given node has $k$ connections to other nodes. This is probably the most important property of networks, and it is behind their classification as exponentially distributed Watts-Strogatz (WS) networks ($p(k)\sim e^{-\alpha k}$) and scale-free Barab\'asi-Albert (BA) networks ($p(k)\sim k^{-\gamma}$). For purely random Erd\"os-Renyi networks, $p(k)$ is a binomial distribution (so also belonging to the exponentially distributed class,
\begin{equation}
p(k)={N \choose k} p^{k}\left(1-p\right)^{N-k}\nonumber
\end{equation}
Sparse networks are those for which the average degree remains finite when $N\rightarrow\infty$, and for real networks, $<k>\ll N$.}

\item{Clustering: The clustering coefficient $c_i$ of a vertex $i$ is given by the ratio between the number $e_i$ of triads -connected subsets of three network nodes- sharing that vertex, and the maximum number of triangles that the vertex could have. Alternatively, this coefficient is the ratio between the number $E_i$ of edges that actually exist between the $k_i$ neighbors of vertex $i$ and the maximum number $k_i(k_i-1)/2$. The clustering coefficient provides a measure of the local connectivity structure of the network. This coefficient usually takes large values in social networks, contrary to what happens in random graphs \cite{reviewAB}. The average cluster coefficient is given by $<c>=\sum_i c_i/N$ and the clustering spectrum by $<c(k)>=\sum_i \delta_{kk_i}c_i/Np(k)$, where $N$ is the number of vertices in the network, and $p(k)$ is the degree distribution defined below. A graph is considered to be small-world if its clustering coefficient is considerably greater than that of a random graph built on the same node set and the average path length is approximately the same as that of the corresponding random graph. One may point out that the clustering coefficient of triangular Erd\"os-Renyi networks, i.e. uncorrelated random graphs, is very high by construction, but, contrary to intuition, it is different from 1 in general.}

\item{ The overlapping index  \cite{gligorausloosAOI} measures the common number of neighbors of  the $i$ and $j$ nodes, i.e. how many triads have a common basis. Moreover, a network transitivity is the probability that two neighbors  of a node have themselves a link between them. In topological terms, it is a measure of the density of triads in a network.}

\end{enumerate}
\subsection{Additional parameters and concepts}
In addition to the ones mentioned above, there are a lot of other parameters and concepts  that measure important properties of the topology of complex networks. A thorough treatment of these can be found in more detailed reviews like the ones cited in this chapter or in monographs like for e.g. that in Ref. \cite{PastorSatorrasRDG}. For brevity, we shall mention just a few of them of particular interest.

\begin{enumerate} \item {Centrality of a node \cite{betweenness}. Many measures are gathered under the label ``Centrality''. The node degree, i.e. its number of contacts to other nodes of the network, is one of them. Another one is the so called betweenness of a vertex $b_i$ or an edge $b_{ij}$, which is the number of shortest paths that pass through the vertex $i$ (edge $(i,j)$), for all the possible pairs of vertices in the network.}

\item{Correlations in networks: The correlations usually found in real networks (i.e. the fact that the degrees of the nodes at the ends of a given vertex are not in general independent) are measured by means of the distribution $p(k\mid k')=k'p(k')/<k>$, representing the conditional probability that an edge that has one node with degree $k$ has a node with degree $k'$ at the other end. In a correlated network, this distribution depends both in $k$ and $k'$, while in an uncorrelated network, it depends only on $k'$. An alternative measure of correlations is given by the average degree of the nearest neighbors of the vertices of degree $k$ \cite{correlations},
\begin{equation}
<k_{nn}>=\sum_k^{'} k'p(k\mid k')
\end{equation}
The network is said to be correlated if this parameter depends on $k$. Other measures are the closeness centrality or the flow-betweenness centrality. When $<k_{nn}>$ increases with $k$ the network is called assortative, while if $<k_{nn}>$ is a decreasing function of $k$, the network is called disassortative. Then, the assortativity coefficient  \cite{assortativity} measures a network property through the Pearson correlation coefficient of the degrees at either ends of an edge. Finally, one must recall that $p(k\mid k')$ and $p(k)$ are not independent, but, due to the conservation of edges, they are related by a degree detailed balance condition \cite{Boguna}
\begin{equation}
kp(k'\mid k)=k'p(k')p(k\mid k')
\end{equation}}
\item {$k$-shells. A $k$-shell of a graph $G$ is a connected subgraph of $G$ in which all vertices have degree at least $k$. Equivalently, it is one of the connected components of the subgraph of $G$ formed by
repeatedly deleting all vertices of degree less than $k$. The $k$-core of a graph $G$ is the $k$-shell with the maximum $k$. The concept of $k$-shells and $k$-cores was introduced to study the clustering structure of social networks and to describe the evolution of random graphs; it has also been applied in bioinformatics and network visualization. In Economics and Finance it has been applied to the corporate ownership network, to the international trade network, and to the network of shareholders \cite{Panos,Garas,Rotundo}.}
%$k$-shells. A $k$-core of a graph $G$ is a maximal connected subgraph of $G$ in which all vertices have degree at least $k$. Equivalently, it is one of the connected components of the subgraph of $G$ formed by repeatedly deleting all vertices of degree less than $k$. 
%The concept of a $k$-core was introduced to study the clustering structure of social networks and to describe the evolution of random graphs; it has also been applied in bioinformatics and network visualization.}
\item {Nestedness index \cite{AraujoPhA389}}. It indicates the likelihood that
a node is linked to the neighbors of the nodes with larger degrees. The mean topological overlap between nodes \cite{Almeida-Neto} has been introduced to quantify nestedness.

\end{enumerate}
\subsection{Ising model on complex networks}
Several classical problems of statistical mechanics have been now studied using complex networks. In particular, the mean-field solution for the average path length and for the distribution of path lengths in small-world networks have been reported by Newman et al. \cite{NewmanMooreWatts}. On the other hand, the mean-field solution of the Ising model \cite{Ising} on a small-world complex network has been contributed by several authors \cite{Gitterman,Barrat,Viana-Lopes,Herrero}, and by Bianconi on a BA network \cite{Bianconi} in the first half of the last decade. Viana-Lopes \cite{Viana-Lopes} solved the 1D Ising model on a small-world network, and Herrero \cite{Herrero} considered the ferromagnetic transition for the Ising model in small-world networks generated by rewiring 2D and 3D lattices.
 %PHYSICAL REVIEW E, VOLUME 65, 066110
Due to its interest for socioeconomic researchers, some of the main results reported in these contributions are recalled below.

The Ising model \cite{Ising}, originally introduced for the study of ferromagnetism, is the simplest paradigm of order-disorder transitions. Undoubtedly it represents one of the major milestones in the development of statistical mechanics of interacting systems and phase transitions, and it is at the basis of a plethora of applications and generalizations reported throughout the 20$^{\mathrm{th}}$ century. This is a discrete model in which spins (agents) with two possible states ($\pm 1$) are placed in the nodes of a graph (normally a lattice) and are allowed to interact with their nearest-neighbours, being probably the simplest model to exhibit a phase transition. The one dimensional problem was solved by Ising himself in 1925 \cite{Ising}, and the exact solution to the 2D problem was reported by Onsager in 1944 \cite{Onsager1944}. The model Hamiltonian for the $N$ spins $s_{i}=\pm 1$ on the nodes of the graph is given by

\begin{equation}
H=-\sum_{i,j}J_{i,j}s_{i}s_{j}-\sum_{i}h_{i}s_{i}
\label{Isinghamiltonian}
\end{equation}
$h_{i}$ being the local external (magnetic) field, and $J_{ij}$ being nonzero only for those pairs of spins connected by a link. If $J_{ij}>0$ then parallel orientations of spins are energetically favored (ferromagnetism), while for $J_{ij}<0$ antiparallel orientations are preferred (antiferromagnetic case). In social physics applications these cases correspond, respectively, to consensus/dissensus-oriented models. For the nearest neighbours 1D problem, eq. (\ref{Isinghamiltonian}) can be rewritten as 

\begin{equation}
H=-J\sum_{i=1}^{N}s_{i}s_{i+1}-\sum_{i=1}^{N}h_{i}s_{i},
\label{Isinghamiltonian1D}
\end{equation}
with $s_{N+1}=s_{1}$ as usual for periodic boundary conditions. The canonical partition function -the normalization factor of the probability density in the phase space of the system's microstates- is straightforwardly calculated from the above equation as:

\begin{eqnarray}
Z(\beta)&=&e^{-\beta H(\sigma)}\nonumber\\
&=&\sum_{s_{1}...s_{N}}e^{\beta h s_{i}}e^{\beta s_{i}s_{i+1}}\nonumber \\
&=&\sum_{s_{1}...s_{N}}\prod_{i=1}^{N}\Delta_{s_{i}s_{i+1}}
\label{Isingpartfunc1D}
\end{eqnarray}
where $\beta^{-1}=k_{B}T$ represents the thermal energy and $\Delta_{s_{i}s_{i+1}}=\exp(\beta h s_{i})\exp(\beta s_{i}s_{i+1})\\
exp(\beta h s_{i+1})$ is the transfer matrix. Thus, using conventional matrix algebra, we can obtain the partition function in terms of the eigenvalues of the matrix $\Delta^{N}$, $\lambda_{i} (i=1,2)$, as:
\begin{eqnarray}
Z(\beta)&=&\mathrm{Tr}\left(\Delta^{N}\right)\nonumber \\
&=&\lambda_{1}^{N}\left[1+\left(\frac{\lambda_{2}}{\lambda_{1}}\right)^{N} \right]
\end{eqnarray}
and the associated Helmholtz free energy per spin (a thermodynamic potential comprising all the relevant thermodynamic information about the system and governing equilibrium and stability at constant temperature and volume), $f(\beta)$, as:
\begin{eqnarray}
-k_{B}Tf(\beta)&=& \lim_{N\rightarrow\infty}\frac{Z(\beta)}{N}\nonumber\\
\label{Isingfreenergy}
&=&\ln\left(e^{\beta J}\cosh(\beta h)+\sqrt{e^{2\beta J}\sinh^2(\beta h)+e^{-2\beta J}}\right)
\end{eqnarray}
Moreover, one can prove (see for example Ref. \cite{LeBellac} for an elegant treatment of the topic) that spin-spin correlations are given in this model by:
\begin{equation}
\left\langle s_i s_j\right\rangle=e^{-r_{ij}/\xi}
\end{equation}
where $\xi=a/\arrowvert\ln \tanh\left(\beta J\right)\arrowvert$, with $a$ being the lattice spacing between spins. It is well-known \cite{LeBellac} that a 1D system exhibits no phase transition in the absence of long-range interactions. In 2D systems this ferromagnetic transition ($J>0$) is registered, and it was Onsager \cite{Onsager1944}, who obtained the partition function for the vanishing external magnetic field case, and Yang \cite{Yang} who calculated the magnetization in the ferromagnetic phase given by

\begin{equation}
M=\left\{{1-\left[\sinh\left(\log(1+\sqrt{2})\frac{T_{c}}{T}\right)\right]^{-4}}\right\}^{1/8}
\end{equation}
Here $k_{B}T_{c}=2J/\log(1+\sqrt{2})\simeq 2.27J$ is the critical temperature where the ferromagnetic (consensus) transition takes place. 
%\marginpar{Marcel: (1.10) verified in several fonts}

Contrarily to their physical homologues, where interactions are usually of limited  range, in social systems long-range connections between agents are very frequently registered. In order to preserve an Ising-based descriptions of these systems, it is necessary to generalize the formalism so as to include the existence of long-range correlations between agents, and that is most conveniently done by means of complex networks. 

The solutions of the 1D Ising model on a small-world network has been reported by Viana-Lopes et al. \cite{Viana-Lopes}, and on a scale-free network by Bianconi \cite{Bianconi}, the latter in the mean field approximation. In the former work, the authors exactly solved a one-dimensional Ising chain with nearest neighbor interactions and random long-range interactions, obtaining a phase transition of a mean-field type. The authors considered a 1D lattice in which the bonds are rewired at random with a probability $p$ in a Watts-Strogatz fashion, and they used a Hamiltonian where they allowed the existence of different short-range (chain, $J$) and long-range ($I$) interactions,
\begin{equation}
H=-J\sum_{i=0}^{N}s_{i}s_{j}-I\sum_{{ij}\in S}s_{i}s_{j} -h\sum_{i=1}^{N-1}s_{i}
\label{IsinghamiltonianSW}
\end{equation}
where the set $S$ includes the $N_{b}=Np$ shortcut pairs of nodes connected by a long-range connection. Even for small values of the shortcut probability $p$, a dramatic increase in the connectivity of the network is registered, which has a deep influence in the thermodynamics of the Ising problem. Particularly, Viana-Lopes et al. \cite{Viana-Lopes} proved that a phase transition exist in this 1D problem with a transition temperature given by
\begin{equation}
t_{J}^{d}(1+2t_{I})=0
\label{IsingSWcrittemp}
\end{equation}
where $d=1/2p$, $t_{J}=\tanh(\beta J)$ and $t_{I}=\tanh(\beta I)$. The authors analyzed the behavior of this temperature in the limits of shortcut bonds stronger than chain bonds ($pI\rightarrow\infty$ for any finite $p$) and of chain bonds much stronger than shortcut bonds ($pI\rightarrow 0$) and obtained:
\begin{eqnarray}
T_{c} &=& \frac{2J}{\ln\left(1/p\ln3\right)} \hspace{2.4 cm}  pI\rightarrow\infty\nonumber\\
T_{c} &=& \frac{2J}{\ln\left\{J/\left[pI\ln\left(J/pI\right)\right]\right\}} \hspace{1 cm}  pI\rightarrow 0
\end{eqnarray}

On the other hand, the Ising model on a BA network has been solved in the mean-field approximation by Bianconi in Ref. \cite{Bianconi}, as previously mentioned. The author showed that the mean-field solution of the Ising model in this type of network can be treated as a Mattis model, a simple solvable model of the spin glass in which Ising spins interact via unfrustrated random exchange interactions \cite{Mattis}. The author considered a BA network of $N$ spins constructed iteratively with the constant addition of new nodes with $m$ connections and a Hamiltonian 
\begin{equation}
H=-J\sum_{i,j}\epsilon_{ij}s_{i}s_{j}-\sum_{i=1}^{N}h_{i}s_{i},
\label{IsinghamiltonianBA}
\end{equation}
Here $\epsilon_{ij}=\langle A_{ij}\rangle=k_{i}k_{j}/2mN$ is the average of the adjacency matrix over many copies of the network \cite{Bianconi}. The mean-field solution of the Hamiltonian for the order parameter $S$ is

\begin{eqnarray}
S &=& \frac{1}{2mN}\sum_{i=1}^{N}k_{i}\langle s_{i}\rangle\nonumber\\
&=&\frac{1}{2mN}\sum_{i=1}^{N}k_{i}\tanh\left[\beta\left(Jk_{i}S+h_{i}\right)\right]
\end{eqnarray}
where one can see that the effective mean-field acting on a spin is determined not only by the external field and the interaction strength, as in conventional Ising model, but also by the connectivity of the network nodes. The above equation resembles that of the Mattis model with the substitution of the quenched random variables $\xi_{i}$ in Ref. \cite{Mattis} by the node degree $k_{i}$. Bianconi was able to prove from the above that the effective critical temperature is given by
\begin{equation}
T_{c}=\frac{mJ}{2}\ln(N),
\end{equation}
i.e. it increases linearly with the interaction $J$ and logarithmically with the number of nodes in the system, in agreement with previously reported numerical simulations (see Ref. \cite{Bianconi} for further details).

%Lambiotte, R. and Ausloos, M., Modeling the evolution of coupled networks, First World congress on Social Simulation (WCSS 2006) e-Proceedings, vol.1, pp. 375- 381  \cite{496RLMAwcss} 

% Ausloos, M. and  Lambiotte, R., Drastic events make evolving networks, \emph{European Physical Journal B}, 57 (2007) 89-94 \cite{520MARLdrastic} 

% Lambiotte, R. and Ausloos, M., Growing network with j-redirection, Europhys. Lett. 77 (2007) 58002 $ (http://arxiv.org/abs/physics/0612148, 35kb) $ \cite{511RLMAj} 
 
Finally, it is worth mentioning that the existence of phase transitions in 1-D systems with long-range interactions has been proved by means of Monte Carlo simulations by Pekalski \cite{Pekalski}, who 
demonstrated that even a small fraction of long-distance shortcuts induces ordering of the system at finite temperatures and provided the dependence of the magnetization and the critical temperature on the concentration of the small world links.

\subsection{A special case: Bipartite networks reductions}

For our present purposes, it must be here emphasized that when discussing interacting economic entities,  it is paramount to discriminate $N$-body correlations that are intrinsic $N$-body interactions from those that merely develop from lower-order interactions, like the 2-body interactions of the Ising model. This issue is directly related to a well-known problem in complex network theory, i.e. the projection of bipartite networks composed of two kinds of nodes, onto unipartite networks, i.e., composed of one kind of node. This property of bipartiteness is a special case of disassortativity. A network is called bipartite if its vertices can be separated into two sets such that edges exist only between vertices of different sets. %Lambiotte, R., Ausloos, M.,  N-body decomposition of bipartite networks,  Phys. Rev. E  72  (2005) 066117 (8 pages)
\cite{467RLMApre72}. 
Bipartite networks are well known in graph theory and operations research, where the  delivery problem from $N$ sources to $M$ sinks is well studied, and finds a first application in Economics to the problem of finding the optimal supply of goods (in the $N$ sources) to accomplish the demand function of the $M$ consumers (the $M$ sinks) \cite{Hotelling}. The model may vary to consider the optimal location and geographical distance among economic activities (the $N$ sources) and the customers (the $M$ sinks). Further recent applications to financial networks relate the set of $N$ companies to the set of their $M$ directors, who are persons, so the two sets are naturally describing two very different categories. This is naturally a bipartite graph, and there is a link among a company and a person if he/she is in the administrative board of the company.
This network can be represented through a matrix $A \in R^{N\times M}$ that is the starting point for the study of ties among companies given by the presence of the same directors in their boards ($AA^T$), or the connections among persons due to belonging to the same boards ($A^TA$)  \cite{garlaschelli,bertone,GRAMDAASS,grassi,croci}.

This formalism and a coarse graining description of bipartite networks from a statistical mechanics approach can be found in 
\cite{467RLMApre72,music1,music2,CPC147.02.40Newman}. Formally, the bipartite structure of e.g., quantities or prices $vs.$ producers may be mapped exactly
on the $vector$ of matrices $\mathcal{M} $ defined by: \begin{equation}
\label{one} \mathcal{M} = [M^1_{a1} , M^2_{a_1 a_2} ,  M^3_{a_1 a_2 a_3}
,...., M^{n_P}_{a_1... a_{n_P}}] \end{equation} where  ${\bf M}^j$ is a
square $n_P^j$ matrix that accounts for all quantities (at some price)  $j$ produced  by producer  $P$.   For example,
$M^1_{a_1}$ and $M^2_{a_1 a_2}$ represent respectively the total number
of goods produced by $a_1$ alone, and the total number of  goods produced  
by the pair ($a_1$, $a_2$).

It is important to point out that the vector of matrices $\mathcal{M}$
describes  the bipartite network without approximation, and that it
reminds of the Liouville distribution in phase space of a Hamiltonian
system. Accordingly, a relevant macroscopic description of the system
relies on a coarse-grained reduction of its internal variables. The
simplest reduced matrix is the single producer matrix: 
\begin{equation}
R^1_{a_1} = M^1_{a1}  +  \sum_{a_2} M^2_{a_1 a_2} + 
\sum_{a_2}\sum_{a_3<a_2} M^3_{a_1 a_2 a_3}+ .... +  \sum_{a_2}....
\sum_{a_j<a_{j-1}} M^j_{a_1 ... a_j}+... 
\end{equation} 
that is a vector whose elements $R^1_{a_j}$ denote the total number of  goods
produced by   $a_j$. The second order matrix:
\begin{equation} R^2_{a_1 a_2} = M^2_{a_1 a_2} +  \sum_{a_3} M^3_{a_1
... a_3}+....+  \sum_{a_3}.... \sum_{a_j<a_{j-1}} M^j_{a_1 ... a_j}+...
\end{equation} Its elements represent the total number of quantities
produced by the pair   ($a_1$, $a_2$). 

Remarkably, this matrix reproduces the usual projection method   and  obviously simplifies the bipartite structure by hiding the effect of higher order interactions.  One may next discriminate between different types of triangles and discuss, e.g. the interplay between producers at the node degree level. Economic directed and weighted networks such as payment networks (see Bougheas and Kirman in this volume) needs to be further explored.

\section{Computational description of complex networks}
The representation of graphs and networks as data structures in computer memory is by now well established, with several comprehensive and efficient implementations openly available \cite{BoostGraph}. The representation determines both the storage requirements and the efficiency of common operations on networks, and must be chosen accordingly. It must be noted that the available formats differ mostly in how edges are stored, as the most appropriate structure for storing data about vertices is relatively independent of the topology.

A particularly simple data structure for storing graphs is the adjacency matrix $A=(a_{ij})$ with as many rows and columns as edges in the graph. Among adjacency matrices, the less complex are those of simple undirected graphs with no edge weights, where the diagonal elements are always zero, $a_{ij}=a_{ji}$ and each element is $1$ if the corresponding edge exists and $0$ otherwise. However, when edge directionality is introduced the symmetry constraint must be abandoned, and if edges have weights they can be used as matrix entries. The main virtue of adjacency matrices, beyond the straightforwardness of their implementation, is the efficiency of adding/removing edges and checking for their existence. On the other hand, they are particularly poorly suited for representing social networks, which tend to be sparse: in a naive implementation of an adjacency matrix for a graph with $N$ vertices, most of the $N^2$ entries will be zero. However, these matrices can still be used as long as they are stored in one of the formats commonly employed for sparse arrays, such as a coordinate list (list of tuples $\left(i,j,a_{ij}\right)$ containing only those for which $a_{ij}\ne 0$) or a dictionary using the $\left(i,j\right)$ coordinates of non-zero elements as keys \cite{sparse}. Alternative, specific formats exist for graphs. For instance, in an adjacency list the set of neighbors is stored along with the rest of the data for each vertex, which keeps vertex addition and removal very efficient while ensuring that storage space scales only linearly with $N$.

It is not often that one finds the opportunity, or even the need, to model a real-world social network in a completely detailed fashion. Although such studies might be useful to assist political decisions, most theoretical approaches are commonly focused on exploring general phenomena, for which such overfitted models would be of little help. Instead, ensembles of random networks are built that capture just the essential features of the real networks. Of interest for social models is the fact that in most generating algorithms the structure of a network is a phenomenon that emerges from its growth dynamics. For instance, maximally random networks, where each of the $N\left(N-1\right)/2$ edges has the same probability ($p$) of being present regardless of the degrees of the vertices it joins, can be created using the classical Gilbert \cite{Gilbert} algorithm, in which edges are added at random to an initially disconnected graph. The results are equivalent to those of the Erd\"os-Renyi \cite{Erdos-Renyi} model: networks that have no loop with lengths much shorter than the size of the graph. These local tree-like loops give rise to the small-world property of purely random graphs, in which the average  distance between nodes grows only proportionally to the logarithm of $N$. Another very important phenomenon observed in these simple models is the emergence of a giant connected component, comprising a finite fraction of all nodes even in the infinite-network limit, when $p>1$.

Both of the aforementioned properties have been observed repeatedly in real-world social networks. One of the most striking examples is a recent comprehensive study of the structure of the Facebook "friend" network, with $721$ million users \cite{Facebook}, $99.91\%$ of them in a single connected component and an average distance between nodes of just $4.71$. However, the limitations of Erd\"os-Renyi for describing real networks are also readily encountered when looking at the local environment of each vertex. The distribution of node clustering coefficients (number of links that exist among the neighbours of a degree-$k$ node, normalized to the maximum number $k\left(k-1\right)/2$ that could exist) or the network clustering coefficients (fraction of the total possible $3$-loops, or triads, actually present in the graph) tends to zero in the infinite-network limit and underestimates observed values by two to three orders of magnitude. This is a reflection of the fact that in real-life processes such as interactions between individuals links are more likely to be formed with other nodes in the local environment as opposed to distant ones, a bias not taken into account at all in simple random models. Another trivial limitation is that the Bernoulli processes used to build these families of graphs result in Poisson-like exponential distributions, while the shape of experimental distributions is often found to be less quickly decaying, and even scale-free.

The first problem is tackled by modified construction algorithms such as the celebrated Watts-Strogatz model \cite{SWN}, that takes a regular network with a ring topology (high clustering coefficient, long mean  distance between nodes) as its starting point and proceeds by replacing its edges with random shortcuts with a uniform probability $p$. The small-world limit is quickly reached even for small values of $p$, whereas the clustering coefficient decays more slowly as $\left(1-p\right)^3$. Hence, Watts-Strogatz networks interpolate between the desirable properties of regular graphs and the Gilbert model. However, their degree distribution is still exponential for large $k$.

In contrast, the Barab\'asi-Albert model \cite{reviewAB} for network growth dynamics achieves a scale-free degree distribution with a $\propto k^{-3}$ fat tail by using a preferential attachment mechanism: the starting point is a small and fully connected network and at each step a new node is added and connected to $m$ of the existing ones with probabilities proportional to their degrees. The very connected nodes present in the final network result in mean distances between nodes even shorter than those in small-world networks. Other functional choices for representing the preference for attachment with well-connected nodes are possible, and as long as they are asymptotically linear they also result in scale-free degree distributions \cite{genBA}, with exponents down to $2$. This limit can be overcome by abandoning pure preferential attachment and introducing more drastic measures, such as merging of vertices \cite{merger}, into the dynamics of the network. It must be mentioned here that, since real studies are always performed on finite networks, and fat tails may only become clear at high values of $k$, deciding on whether a real network is scale-free or not can be mathematically and computationally challenging; in fact, different groups working on the same data have occasionally reached opposite conclusions \cite{exponentialorscalefree}.

Still, none of these models offer the researcher the possibility to computationally build ensembles of networks with a known degree distribution, possibly taken from experiment. Fortunately, it is straightforward to generalize, for instance, the idea behind the Erd\"os-Renyi model to sample graphs uniformly from the ensemble formed by all those with the desired degree distribution \cite{generalized}. The key feature of such models is the fact that, by construction, the degrees of the nodes at either end of the same link are not correlated. More specifically, their joint degree distribution $p\left(k,k^{\prime}\right)$ can be factorized as $p_e\left(k\right)p_e\left(k^{\prime}\right)$; here, $p_e\left(k\right)$ is the degree distribution of an end of a randomly chosen edge, related to the node degree distribution $P\left(k\right)$ by $p_e\left(k\right)=p\left(k\right)/k$.  When their degree variance is finite, these uncorrelated networks are also locally tree-like, but their clustering coefficients in finite cases approximate those of real networks much better than the ones derived from the Gilbert model. Moreover, it is still possible to derive a general condition for the emergence of a giant connected component in this setting, the Molloy-Reed criterion \cite{molloy1,molloy2}, $\left\langle k^2 \right\rangle> 2\left\langle k\right\rangle$. This criterion is met by scale-free networks, whose second moment diverges in the infinite limit, which explains their extreme resilience.

Clearly, not even the most general uncorrelated networks can capture all the varieties of real-world structures. Even though the joint distribution $P\left(k,k^{\prime}\right)$ is seldom available, the Pearson correlation coefficient between $k$ and $k^{\prime}$, known in this context as assortativity ($r$) can be used \cite{assortativity} to discriminate between correlated and uncorrelated networks. Preferential-attachment algorithms such as Barab\'asi-Albert give rise to assortative ($r>0$) networks, while electrical grids, for instance, are know to be dissortative ($r<0$) \cite{exponentialorscalefree}.

A computational study of a network in the time domain will typically start with a network in an initial condition that can be empty, comprehensively sampled from real-world data or generated using one of the algorithms described above and incorporating only the necessary information. It will then proceed through a set of discrete time steps until the desired convergence is achieved. At each step, both the structure of the network and any node or edge variables may change in response to external fields, internal phenomena or interactions through the network. Several points make this kind of simulation very different from models based purely on differential equations, with a longer tradition in physics and economics. First, the computational demand of agent-based models can be formidable when compared with more aggregate treatments of similar systems; thus, automation and parallelization are critical. Even when the network dynamics can be implemented synchronously, with changes only depending on data from previous steps, to avoid race conditions, scaling can be hindered by the need to exchange large amount of data during updates between steps. Moreover, judicious choice of what aggregate variables to track to characterize the evolution of the network \cite{496RLMAwcss,520MARLdrastic} is of crucial importance to separate signal from noise. In addition to a set of basic descriptors (number of nodes and links, weights and directions if applicable, and so on) and degree and clustering distributions, more problem-specific metrics such as assortativity, betweenness centralities \cite{betweenness}, and overlap \cite{gligorausloosAOI} and the nestedness indices may be necessary. Finally, given the very specialized nature of network visualization, storage of data should be done in standard formats (HDF5, NetCDF) and the network itself made available in well-documented formats such as GraphML and GML to retain freedom to look at the results from different tools and environments. Third, judicious choice of what aggregate variables to track is of crucial importance to separate signal from noise. Open, scalable, general-purpose visualization tools \cite{pajek,gephi,python} are currently available to ease post-processing work.

\section{The economy as a complex adaptive system: Complex-network-based market models}
In the last decades, there has been an increasing amount of effort to conceive the economy as an evolving complex system (see for example Ref.\cite{AndersonArrowPines,Tesfatsion,Kirman1997,BarkleyRosser1999,ColanderHoltRosser,BrianArthur2006,Foster2005,MartinSunley,MJackson}), meaning that it is a dynamic network of interactions between similar, connected agents which self-organize in order to adapt to a changing environment and maintain the organization of the macrostructure. Many of these approaches employ the complex network formalism -either as a theoretical or a computational tool- for analyzing different perspectives of the economic realm: (i) economics formalism itself, (ii) finance, and (iii) social networks applied to economics. Below, some of the main contributions in the field are reviewed.

Theoretical and computational agent-based models of economic interactions are being progressively more used by the scientific community to describe the emergent phenomena and collective behavior in economic activities arising from the interaction and coordination of economic agents \cite{NamatameKA}. The complex-network formalism is essential for most of these descriptions. This kind of approach has been used in several problems such as international trade \cite{Bhattacharya}, finance \cite{Caldarellinature,daCruz}, globalization \cite{Kali,Rodrigue,Xiang_Lia,Garlaschelli2, Fagiolo} and so forth). Specifically, Schweitzer et al. \cite{Fagiolo} analyzed economic networks as tools for understanding contemporary global economy. These authors considered the ability of these objects to model the complexity of the interaction patterns emerging from the incentives and information behind agents' behavior from which metastabilities, system crashes, and emergent structures arise. A detailed characterization of these phenomena requires, according to the authors, ``a combination of time-series analysis, complexity theory, and simulation with by game theory, and graph and matrix theories".

%\bibitem{Fagiolo}ECONOMIC NETWORKS: WHAT DO WE KNOW AND WHAT DO WE NEED TO KNOW? FRANK SCHWEITZER, GIORGIO FAGIOLO, DIDIER SORNETTE, FERNANDO VEGA-REDONDO, DOUGLAS R. WHITE, Volume: 12, Number: 04n05 (August & October 2009)

\subsection{Economics: Gross Domestic Product and other macroeconomic indicators}

Many contributions have been reported that consider the application of complex networks to the analysis of gross domestic product either on a national basis or from a global perspective. Fujiwara et al. \cite{FujiwaraAoyama} considered the production network formed by a million firms and millions of supplier-customer links, showing ``in the empirical analysis scale-free degree distribution, disassortativity, correlation of degree to firm-size, and community structure having sectoral and regional modules".
%\bibitem{FujiwaraAoyama} Y. Fujiwara, H. Aoyama. (2010) Large-scale structure of a nation-wide production network. The European Physical Journal B 77:4, 565-580. Online publication date: 1-Oct-2010
Moreover, Lee et al. \cite{LeeK-M} considered the influence of the global economic network topology to the spreading of economic crises, showing, by means of network dynamics, that its connectivity in the global network conditions the role of the nation in the crisis propagation together with its macroeconomic indicators.
%\bibitem{LeeK-M} Lee K-M, Yang J-S, Kim G, Lee J, Goh K-I, et al. (2011) Impact of the Topology of Global Macroeconomic Network on the Spreading of Economic Crises. PLoS ONE 6(3): e18443. doi:10.1371/journal.pone.0018443

The problem of measuring the degree of globalization of the economy is also an outstanding question. One method is to search for a measure of the clustering features through the notion of economic distance. According to Miskiewicz and Ausloos, it is possible \cite{475NetMisk} to develop a distance and graph analysis for doing so. This has been performed on the  GDP of G7 countries over 1950-2003. In fact, several (4) different distance functions can be used and the results compared. Moreover, the graph method takes two forms in  \cite{475NetMisk}, i.e.  (i)  a unidirectional  or (ii) a bidirectional chain. In brief, the (linear) network is allowed to grow, accumulating distances,  in one or two directions.  
	
Defining the percolation transition threshold,  as the distance value at which all countries are connected to the network, it has been found that the correlations
between GDP yearly fluctuations  \cite{475NetMisk,512MGMAepjb57}
achieve their highest value in 1990. Hence, the globalization of the world economy
was seen to disappear as early as 2005 in publications, and is so confirmed nowadays
by economists.
%network, it has been found that the correlations achieve their highest value in 1990. Whence, the globalization of the world economy  was seen to disappear as early as 2005 in publications, and is so confirmed nowadays by economists \textcolor{red}{The reviewer asks: According to which measure has globalization disappeared? You also need to provide references if you refer to ‘publications’}.
		
  %M. Gligor and Ausloos, M., Cluster structure of EU-15 countries derived from the correlation matrix  analysis of macroeconomic index fluctuations, Eur. Phys. J B 57 (2007) 139- 146  
Macroeconomic indicators other than GDP correlations can be used  for such globalization/ infinite cluster search. In  \cite{512MGMAepjb57}, 11 of them were investigated for the 15 original EU countries, - the data taken between 1995 and 2004.  Moreover, besides the Correlation Matrix Eigensystem Analysis, a Bipartite Factor Graph Analysis is of interest for confirming the existence of stable ``economic" communities. It has been interestingly found that strongly correlated countries, with respect to these 11 macroeconomic indicators fluctuations, can be partitioned into clusters, mainly based on geographic grounds.
	 
The Moving Average Minimal Length Path algorithm has allowed a decoupling of the fluctuations \marginpar{Marcel, a ref. would be useful here}. Whence a Hamiltonian representation can be formulated, given by a factor graph. Practically, that means that the Hamiltonian-Liouville machinery could be used thereafter. %\marginpar{The reviewer suggests replacing ``thereafter, even going toward thermodynamics" with ``applying methods used for the study of thermodynamics". In my opinion it changes the meaning of your sentence, right?}. 
In fact, the Hamiltonian plays the role of a cost function.
	 
It is somehow evident that markets can not be instantaneously correlated. %\textcolor{red}{The reviewer asks : What do you mean by ‘intellectual’ and why should there be this type of evidence?)}
%3. Ausloos, M. and Lambiotte, R., Clusters or networks of economies? A macroeconomy study through GDP fluctuation correlations, Physica A 382 (2007) 16-21 \cite{514MARLPhA382}$ (http://arxiv.org/pdf/physics/0607139, 266kb) $ doi:10.1016/j.physa.2007.02.005   
In \cite{514MARLPhA382}, forward and backward correlations, and distances, between GDP fluctuations were calculated for 23 developed countries. In this study, the network links were not only weighted but were also directed due to the arrow of time. Filtering the time-delayed correlations by  successively removing the least correlated links, an evolution of the 23 countries network can be visualized. In so doing, this percolation idea-based method reveals the emergence of connections,  but  interestingly also of leaders and followers.

%4. M. Gligor and Ausloos, M., "Convergence and cluster structures in EU area according to fluctuations in macroeconomic indices", Journal of Economic Integration  23 (2008)  297-330; \cite{537} 
It is also relevant to know what time window has to be used when averaging properties, both in micro- and macro-economic considerations. In \cite{537}, several statistical distances between countries were calculated for various moving time windows.
	Up to 4 macroeconomic indicators were investigated: GDP,  GDP/capita (GDPpc), Consumption, and Investments. %The order of magnitude of the relevant time windows has been identified: a 7-year time window is  necessarily to be used for coherence.
	In using a so called  optimal one, some  empirical evidence  has been presented to indicate economic aspects of globalization through such indicators. The hierarchical organization of countries and  their relative movement inside the hierarchy  have been described. %The  Moving Average Minimal Length Path algorithm allowed to search for a cluster  structure within the EU network.
	  
On the practical side, there are policy implications concerning the economic clusters arising in the presence of Marshallian externalities. Relationships between trade barriers, R\&D incentives and growth were identified. It is recommended that  they should  be accounted for designing a cluster-promotion policy \cite{514MARLPhA382}. Not only it should be admitted that fluctuations in macroeconomic indicators result from nearby node policies, but the effect of policy changes are not immediate. Therefore, some time averaging is necessary in order to find out, how long it takes before some policy affects some country economy and its neighboring or its connected countries.
%5. M. Gligor and Ausloos, M., Clusters in weighted macroeconomic networks : the EU case. Introducing the overlapping index of  GDP/capita fluctuation correlations,    \emph{European Physical Journal B} 63 (2008) 533-539  \cite{539} 
As in \cite{537}, an investigation of the weighted fully connected network of the  25 countries (nodes) forming the European Union in 2005 was presented  \cite{539} to study such time parameter effects. 
	 The links were taken to be proportional to the degree of similarity between the macroeconomic fluctuations and the GDP/capita  (GDPpc) annual rates of growth between 1990 and 2005, measured by the ``coefficients of determination"  \cite{539}.  It has been found that the effect of the time window size for averaging  and finding some robust correlations was a 7-years time window; this time interval leads to coherence in the data analysis and subsequent interpretations. A calculation of the  ``overlap index"  \cite{gligorausloosAOI} reveals the emergence and stability of a ``hierarchy" among EU countries   \cite{537,539}.
	
%6.  Ausloos, M. and M. Gligor, Cluster Expansion Method for Evolving Weighted Networks Having Vector-like Nodes,  Acta Phys. Pol. A 114 (2008) 491-499  \cite{544} 
 The Cluster Variation Method for weighted bipartite networks, discussed in Sect. 2,  was applied in  \cite{544}. The method allows to decompose (or expand) a ``Hamiltonian" through a finite number of components, serving to define variable clusters in a given  (fully connected) network.  In the case studied in \cite{544}, 
	 the  network  was built from data representing correlations between  4 macro-economic features:  Gross Domestic Product (GDP), Final Consumption Expenditure (FCE), Gross Capital Formation (GCF) and Net Exports. Two interesting features were deduced from such an analysis: (i)   the {\it minimal entropy clustering scheme} is obtained from a coupling $necessarily$ including GDP and FCE; 
	(ii)  the {\it maximum entropy} corresponds to a cluster which does $not$ $explicitly$ $include$ the GDP.
	
%7. F. O. Redelico, A.N. Proto, Ausloos, M., Hierarchical structures in the Gross Domestic Product per capita fluctuation in Latin American countries,   Physica A 388 (2009) 3527-3535   \cite{551} 
A new methodological framework for investigating macroeconomic time series has been introduced in \cite{551}, based on a non-linear correlation coefficient.   Features related directly to the Latin American, rather than EU, countries  have been described, i.e. those  Latin American countries where Spanish and Portuguese languages prevail. Again,  clusters,  in the  networks, and emergence of  a "hierarchy"  have been  identified  through  the "Average Overlap Index" \cite{gligorausloosAOI} hierarchy scheme.    

A principal component analysis has further been applied in order to observe and corroborate whether a country clustering structure truly exists \cite{551}, and the results confirmed previously expected results. %\textcolor{red}{The reviewer asks "What do you mean by “common sense”?"}

\subsection{Finance: market correlations and concentration}

In this section some literature concerning
the application of complex networks to financial problems will be reviewed,
without any particular attention being paid to the problem of
banking risk, which is treated in other chapter of this book by
Bougheas and Kirman \cite{BougheasKirmanChapter}.%(see Chapter \colorbox{yellow}{XX}).  
Besides using a network structure, risk is widespread modelled through correlations.
However, correlations wholly describe risk distribution function only if the phenomena under observation are Gaussian, since they correspond to the second order moment of the joint probability distribution. Higher order models could be relevant in more general situations, giving rise to deviations from the Gaussian distribution. This notwithstanding, it is worth mentioning that in option pricing, although there is evidence of deviations from the Gaussian hypothesis \cite{options}, most models used by practitioners predominantly rely on it and thus give rise to self-fulfilling prophecies \cite{levy}.
There is a plethora of quantitative and econometric models to analyse option pricing, while the complex-network perspective for understanding the relative distance among systems and defining clusters is much more recent.
In relation to the complex-network approach it has been remarked that filtering financial data is relevant for introducing a measure of distance based on stock market indices \cite{Mantegna1999,pozzi}. Subsequent interdisciplinary studies show how to extract relevant information through methods first proposed in graph theory, namely Maximum Spanning tree and the Planar Maximally Filtered Graph also under dynamical adjournment \marginpar{Giulia: adjournment$\rightarrow$ delay?}of data \cite{pozzi,calda,pozzi2}.
 
Yet, the correlation matrix is not the only instrument available for understanding systemic risk due to the connections among companies.
Both the cross-shareholding matrix and the board of directors describe ties among companies, that may cause cascade spreading of financial crises, and raise the question of who is controlling the controller, or having a relative size on markets sufficient to pass the critical threshold of market concentration, triggering the antitrust measures \cite{GRAMDAASS,grassi,croci,GRAMDQQ,GRAMDJEIC,DErrico,GRNATO,Rotundo}.
The development of algorithms for detecting market leaders and related optimization problems is studied also using operation research methods \cite{salvemini}, thus confirming the interdisciplinary nature of the field.
Centrality measures on networks are applied for their purpose of evidencing leaders, and consider antitrust policies for the markets. Moreover, due to the globalization of trading, the default or crisis of companies in one country may spread worldwide \cite{Panos,Garas,Vitali,Rotundo}.

As recently claimed by  Lux and Westerhoff, \cite{LuxWesterhoff} "economic theory failed to envisage even the possibility of a financial crisis like the present one. A new foundation is needed that takes into account the interplay between heterogeneous agents." This foundation can be found in the field of complex networks, as stated by Catanzaro and Buchanan in the same issue \cite{Catanzaro}. Several other papers in the same special issue of Nature Physics contribute different applications of complex networks to the field of finance \cite{Caldarellinature,Galbiati}
%\bibitem{LuxWesterhoff}Thomas Lux and Frank Westerhoff, Nature Physics 2013 March 2013 Volume 9 No 3 pp119-197 
%bibitem{Catanzaro} Network opportunity - pp121 – 122 Michele Catanzaro and Mark Buchanan doi:10.1038/nphys2570
%\bibitem{Caldarellinature} Guido Caldarelli, Alessandro Chessa, Andrea Gabrielli, Fabio Pammolli and Michelangelo Puliga, Reconstructing a credit network Nature Physics 9 No 3 pp119-197
%\bibitem{Galbiati}The power to control - pp126 – 128 Marco Galbiati, Danilo Delpini & Stefano Battiston doi:10.1038/nphys2581

Pursuing this new foundation, a great number of results have been reported in the recent past concerning the application of complex networks formalism to the analyses of financial problems. Probably, these new topological objects have not been applied to any other field of economics to a larger extent. Sampling the work in this area we will mention the works of Onnela, Saram\"aki, Kaski and coworkers \cite{Onnela1, Onnela2, Onnela3, Onnela4, Onnela5, Onnela6,Onnela7,Onnela8,Onnela9} for an extensive aplication of dynamic asset trees to financial markets, as well as to clustering, communities  and correlations in these systems.

Oatley et al. \cite{Oatley} analyzed the political economy of global financial network using a network model, and Caldarelli et al. \cite{Caldarellinature} also employed this formalism and statistical mechanics for reconstructing a financial network even from partial sets of information. On the other hand, da Cruz et al. \cite{daCruz} applied non-equilibrium statistical physics to a system of economic agents obeying the Merton-Vasicek model for current banking regulation and forming a network of trades by means of the exchange of an ``economic energy". The authors analyzed the propagation of insolvency (i.e. the falling of an agent below a minimum capital level) in this network and were able to prove that the avalanche sizes are governed by power-law distributions whose exponents are related to the minimum capital level. Avalanches have been proved to occur also due to behavioral aspects like the blindness to small changes in the worldwide
network of stock markets \cite{JVA}. Finally, we will mention the work of Bonanno et al. \cite{Bonanno} who considered correlation-based networks of financial equities.
%\bibitem{Oatley}Thomas Oatley, W. Kindred Winecoff, Andrew Pennock, Sarah Bauerle Danzman. (2013) The Political Economy of Global Finance: A Network Model. Perspectives on Politics 11:01, 133-153. Online publication date: 1-Mar-2013. 
%\bibitem{Caldarellinature} Guido Caldarelli, Alessandro Chessa, Andrea Gabrielli, Fabio Pammolli and Michelangelo Puliga, Reconstructing a credit network Nature Physics 9 No 3 pp119-197
%\bibitem{daCruz}J.P. da Cruz and P.G. Lind The dynamics of ﬁnancial stability in complex networks \emph{European Physical Journal B} (2012) 85: 256
%\bibitem{Bonanno} G. Bonanno, G. Caldarelli, F. Lillo, S. Micciché, N. Vandewalle, R. N. Mantegna, The European Physical Journal B - Condensed Matter and Complex Systems March 2004, Volume 38, Issue 2, pp 363-371. We review the recent approach of correlation based networks of financial equities. We investigate portfolio of stocks at different time horizons, financial indices and volatility time series and we show that meaningful economic information can be extracted from noise dressed correlation matrices. We show that the method can be used to falsify widespread market models by directly comparing the topological properties of networks of real and artificial markets.
\subsection{Tax evasion}
The network approach has also been applied to the study of tax evasion by Westerhoff et al.\cite{Westerhoff08PhA387}. The authors use the 
%1. Zaklan, G., Lima, W. and Westerhoff, F. (2008): Controlling tax evasion fluctuations. Physica A, Vol. 387, 5857-5861.\cite{Westerhoff08PhA387}
standard two-dimensional Ising model treated in Section 2 to analyze the effect of the structure of the underlying network of taxpayers on the the time evolution of tax evasion in the absence of measures of control. Furthermore, it is shown that ``even a minimal enforcement level may help to alleviate this problem substantially". The number of applications of the Ising model is thus augmented suggesting an enforcement mechanism to policy-makers for reducing tax evasion. 

% 2. Zaklan, G., Westerhoff, F. and Stauffer, D. (2009): Analysing tax evasion dynamics via the Ising model. Journal of Economics Interaction and Coordination, Vol. 4, 1-14. \cite{ZWesterhoffS09JEIC4} 

 %A model of tax evasion is   developed based on the Ising model. We augment the model using an appropriate enforcement mechanism that may allow policy makers to curb tax evasion.
Moreover Zaklan et al. in \cite{ZWesterhoffS09JEIC4} allow tax evaders to be randomly subjected to audits, assuming that if they get caught they behave honestly during certain time. Considering different combinations of parameters, they proved that using punishment as an enforcement mechanism can effectively control tax evasion.
 
\subsection{Business and spreading of innovations}

Beyond the numerous applications of complex networks reviewed in the previous section, these networks have been also applied in other fields of Economics. This formalism is particularly useful for describing the introduction of innovations in markets and/or regions, and has been thoroughly used for that purpose. This methodology has also been used for analyzing business networks. Here some contributions devoted to the study of profit optimization under technological renewal are reviewed, along with dynamic models of oligopoly with R\&D externalities on networks, topics on upstream/downstream R\&D networks and welfare and the spreading of products in markets or co-workers networks.

Diffusion in complex social networks has been considered by several authors. D. L\'opez-Pintado \cite{Dunia1} analyzed the spreading of a given behavior in a population by considering mutual neighbor influence in a network of interacting agents by means of a simple diffusion rule. At a mean-field level, she obtained a threshold for the spreading rate for propagation and persistence in populations, This threshold depends on the connectivity distribution of the underlying social network as well as on the selected diffusion rule. More recently \cite{Dunia2}, the same author considered the spread of free-riding behavior in social networks introducing a model for a social network with free-riding incentives, where agents are allowed to decide whether or not to contribute to the provision of a given local public good. By means of equilibria analysis of the induced game, the author reported the influence of the degree distribution of the underlying network in the fraction of free-riders. Moreover, L\'opez-Pintado and Watts \cite{Dunia3} addressed the problem of the collective behavior of individuals facing a a binary decision under the influence of their social network. The authors reported both the equilibrium and non-equilibrium properties of the collective dynamics and a response function under global and anonymous interactions.
%\bibitem{Dunia1}D. L\'opez-Pintado, Diffusion in Complex Social Networks, Games and Economic Behavior, Games and Economic Behavior, (2008) Vol. 62, 573-590.
%\bibitem{Dunia2}  D. L\'opez-Pintado, The Spread of Free-Riding Behavior in a Social Network Eastern Economic Journal (2008) 34, 464–479. doi:10.1057/eej.2008.30
%\bibitem{Dunia3} Dunia López-Pintado, Duncan J. Watts Social Influence, Binary Decisions and Collective Dynamics,Rationality and Society November 2008 vol. 20 no. 4 399-443

Concerning business structure the work of Semitiel-Garc\'ia and Noguera Men\'endez \cite{Semitiel} is relevant, who, using network theory and social network analysis analyzed the influence of inter-industrial structures and the location of economic sectors, on the diffusion of knowledge and innovation. Specifically, they studied the structure and dynamics of the Spanish Input–output system over a thirty-five-year period. 
%\bibitem{Semitiel} María Semitiel-García, Pedro Noguera-Méndez. (2012) The structure of inter-industry systems and the diffusion of innovations: The case of Spain. Technological Forecasting and Social Change 79:8, 1548-1567. 

Business networks have also been considered by Souma et al. \cite{Souma}, who categorized them into bipartite networks, showing the possibility that business networks will fall into the scale-free category. By means of a one-mode reduction the authors were able to approximately calculate the clustering coefficient and the averaged path length for bipartite networks. These quantities were calculated for networks of banks and companies before/after a bank merger, and they reported quantitative evidence that banks merging increases the cliquishness of companies, and decreases the path length between two companies.
%Complex networks and economics \bibitem{Souma}Wataru Souma, Yoshi Fujiwara, Hideaki Aoyama, Physica A: Statistical Mechanics and its Applications Volume 324, Issues 1–2, 1 June 2003, Pages 396–401

In \cite{RotundoCerquetiPuu,CRITOR,RotundoCerquetiAMS}  Cerqueti and Rotundo developed models
for a set of firms producing a single commodity
dealing with the optimal time for the renewal of the technology.
Such models consider the aggregate outcome.
Eventually, the presence of a hierarchical network organization among firms allows
the leader company to propose a financial strategy, but the proposal is followed by the firms at the peripheral of the network
with a certain probability only.
Depending on the connection level among the companies conditions are obtained for the best strategies
to optimize the profit of a district when a technological renewal takes place.
The papers refer to empirical results drawn on the most large databases CENSUS
and COMPUSTAT for shaping the density of companies and the studies are well
suitable for the development of policies for industrial districts \cite{amaral1,amaral2,axtell}
%In \cite{RotundoCerquetiPuu,RotundoCerquetiAMS}  Cerqueti and Rotundo develop a model for a set of firms producing a single commodity
%dealing with the optimal time for the renewal of the technology and in presence of a hierarchical network organization among firms,
%where the leader company proposes a financial strategy, but it is followed by the firms at the peripheral of the network
%with a certain probability. Depending on the connection level among the companies conditions are obtained for the best strategies
%to optimize the profit of a district when a technological renewal takes place. The two papers refer to empirical results drawn on the most large databases CENSUS and COMPUSTAT for shaping the density of companies and the studies are well suitable for the development of policies for industrial districts \cite{amaral1,amaral2,axtell}.

With respect to the use of complex networks in oligopoly analysis, in a series of papers, Bischi and Lamantia \cite{BischiLamanitaI12} considered the possibility of reducing the cost of knowledge gain for firms through sharing R\&D as well as through investments in R\&D cost-reducing activities. These researchers introduced a two-stage oligopoly game for which they analyzed the existence and stability of equilibria in a given network divided into sub-networks. In pursuing such considerations, the authors considered, in the framework of the two-stage oligopoly game, the influence of the degree of collaboration and spillovers on profits, social welfare and overall efficiency \cite{BischiLamanitaI12}. Analytical results are provided for two relevant cases performing numerical experiments and emphasizing the role of the level of connectivity (i.e. the collaboration attitude) inside networks. The effects of unintentional knowledge spillovers inside each network and between competing networks are also considered in \cite{BischiLamanitaII12}.  
%2. Gian Italo Bischi, Fabio Lamantia, A dynamic model of oligopoly with R\&D externalities along networks. Part II., Mathematics and Computers in Simulation xxx (2012) xxx�xxx \cite{BischiLamanitaII12}

The endogenous formation of $upstream$ R\&D networks have been studied in a vertically related industry and the welfare implications thereof by Kesavayuth et al. \cite{KManasakisZ12JEMS}. The authors reported that in the situation where $upstream$ firms fix prices, the complete network of firms reaches an equilibrium. In contrast, if upstream firms set quantities, a complete network arise only for sufficiently low R\&D spillovers between the firms If these R\&D spillovers are sufficiently high, a partial network arises. Hence, socially optimal equilibrium networks are only reached if upstream firms set prices, and the actual behavior of upstream firms must be taken into account when designing technology policy, and not only the size intra-network R\&D spillovers \cite{KManasakisZ12JEMS}.
%These results affect the social optimality of network formation: when upstream firms set prices, the equilibrium network is socially optimal, while a conflict between equilibrium and socially optimal networks is likely to occur when upstream firms set quantities. Thus, the mode of the ' behavior (prices versus quantities) is as important for designing technology policy as the size of .
%Dusanee Kesavayuth, Constantine Manasakis, and Vasileios Zikos, Upstream R\&D Networks and Welfare, submitted to Journal of Economics and Management Strategy 
 %2. Downstream Research Joint Venture with upstream market power: 
The downstream firms’ incentives in a vertically organized industry have also been examined in \cite{ManasakisPZ12SJE}, where the authors analyzed how and when 
to invest in cost-reducing R\&D, and to form a Research Joint Venture (RJV)%, under two alternative structures of input supply: exclusive i) vertical relations and ii) a single supplier. The authors proved that under non-cooperative investment of downstream firms and low spillovers R\&D investments are higher under a single supplier than under competing vertical chains, contrary to the "hold-up" argument. In this contribution it was also demonstrated that incentives of downstream firms to form an RJV are also stronger in the former case rather than in the latter. 
The authors identified conditions for an RJV to be beneficial to society and discussed integrated innovation and competition policies.
%Conditions under which an RJV is beneficial for society have been identified and Integrated innovation and competition policies discussed by the authors \cite{ManasakisPZ12SJE}.

%Constantine Manasakis, Emmanuel Petrakis, and Vasileios Zikos,  Downstream Research Joint Venture with upstream market power,  submitted to  the Southern Journal of Economics

%2. Dal Forno, A., Merlone, U., (2007). �The evolution of coworkers networks: An experimental and computational approach�, in Edmonds, B., Hern�ndez Iglesias, C., and Troitzsch, K.G., (Eds.), Social Simulation: Technologies, Advances and New Discoveries, Information Science Reference, Hershey (PA), pp. 280-293. \cite{2DalFornoMerlone2007}

Dal Forno and Merlone \cite{2DalFornoMerlone2007} have considered a network of individuals supposing that they could propose and successfully implement their best project. Important elements in the network are: (i) mutual knowledge, (ii) agent coordination in choosing the project to implement,  (iii)  the number of leaders,  and (iv) their location. Leaders increase the social network of other agents making possible projects otherwise impossible; at the same time, they are crucial in setting the pace of a balanced expansion of the social matrix. According to evidence, leaders are not those with the greatest number of connections. The presence of leaders provides a solution to the selection problem when there are multiple equilibria.

Finally, Pombo et al. \cite{pombo} presented evidence of the existence of imitative behaviour among family practitioners in Galicia (Spain), and they used complex network theory and the Ising model (see Sect. 2) in order to describe the entry of new drugs in the market, treating doctors as spins (nodes) in a Watts-Strogatz network. Related to this, one could mention research work done on self-citations of coauthors as defining their research field flexibility, curiosity and in some sense creativity  
 \cite{Ausloos08,Iina06,Iina07}. A combination of such investigations might not only lead to new methods for detecting scientists field mobility, but also indicate pertinent features on new ideas related to the evolution of the production of new goods.

\section{Regional trade, mobility and development}	

Regional trade is another field of economics that has been extensively treated within the framework of complex network formalism. Specifically, a lot of effort has been devoted to the description of the structure \cite{Bhattacharya,Rodrigue,Kali,Garlaschelli2,Fronczak,Fagiolo_Reyes}, communities \cite{Barigozzi} and dynamics \cite{Xiang_Lia} of the global trade network. 
%\bibitem{Xiang_Lia} Xiang Lia, Yu Ying Jin, Guanrong Chen, Complexity and synchronization of the World trade Web, Physica A: Statistical Mechanics and its Applications Volume 328, Issues 1–2, 1 October 2003, Pages 287–296
%\bibitem{Garlaschelli} Diego Garlaschelli, Maria I. Loffredo, Structure and evolution of the world trade network, Physica A 355 (2005) 138–144
%\bibitem{Fagiolo_Reyes}Giorgio Fagiolo, Javier Reyes, Stefano Schiavo, On the topological properties of the world trade web: A weighted, Physica A 387 (2008) 3868–3873. 
%\bibitem{Bhattacharya} Bhattacharya, K., Mukherjee, G., Saramaki, J., Kaski, K. and Manna, S. S. (2008) The International Trade Network, in Econophysics of Markets and Business Networks, New Economic Windows Serie, Springer, pp. 139-147.
%\bibitem{Kali} Raja Kali and Javier Reyes, The Architecture of Globalization: A Network Approach to International Economic Integration.
%\bibitem{Rodrigue}Transportation, Globalization and International Trade THIRD EDITION Jean-Paul Rodrigue (2013), New York: Routledge, 416 pages. ISBN 978-0-415-82254-1
Specifically, in the recent past Fronczak and Fronczak \cite{Fronczak} reported a statistical mechanics study of the international trade network showing that this network is a maximally random weighted network, and that the product of the GDP's of the trading countries is the only characterizing factor of the directed connections associated to bilateral trade volumes. Moreover, Reyes et al. \cite{ReyesWooster} considered the bilateral trade data from the networks perspective, concluding that there seems to be a cyclical pattern in the regional trade agreement formation on the community structure of the world trade network. From this perspective, the pattern of international integration followed by East Asian countries and its comparison with the Latin American performance has been also reported recently \cite{ReyesSchiavo}.

In this subsection we report contributions that, although not truly devoted to network analysis, they depend on the existence of a (fully connected) network.

Among the empirical and quantitative studies, Paas and Schlitte \cite{PaasS08ijrs7} studied the regional income inequality and convergence process in the EU-25. %. Scienze Regionali. Italian Journal of Regional Science, Vol. 7, No. 2, pp.29-49
Paas and Vahi \cite{TPaasTV012} considered the contribution of innovation to regional disparities and convergence in Europe using empirical GDPpc and innovation indicators of the EU-27 NUTS2 the regions. Using principal components factor analysis, three composite indicators of regional innovation capacity were extracted, showing that ca. 60\% of variability of regional GDP per capita is associated to regional innovation performance. Regional innovations are seen to promote the increase of inter-regional differences in the short-run. Consequently, further policy interventions beyond innovation activities should be effectively implemented.  In this respect see also the related  work by Gligor and Ausloos, already mentioned \cite{537,539,544}, on globalization and hierarchical structures in EU, as well as the need of considering appropriate  forward and backward correlations within appropriate time intervals \cite{475NetMisk,514MARLPhA382}. 
%  Integration of  ethnically diverse societies as a source for future economic growth  \cite{TPaasVH012} :  

It is common understanding that international mobility of people and workers are increasing globally. According to Paas and Halapuu \cite{TPaasVH012}, an ethnically and culturally diverse population is expected to create greater variability in the demand for goods and services as well as in the supply of labour through different skills and business cultures favoring new business activities and future economic growth. The authors state that ``although not all immigrants are well-educated and highly-skilled to provide a sufficiently innovative and creative labour force, national economic policies should create conditions that support the integration of ethnic diversity in order to create stable and peaceful environment for economic and political development". Paas and Halapuu aim at clarifying the possible determinants of peoples' attitudes towards immigrants depending on their personal characteristics, as well as attitudes towards households socio-economic stability and a country's institutions. The study's overwhelming aim is to provide empirical evidence-based reasons for policy proposals that, through integration of ethnically diverse societies, creates a favorable climate supporting economic growth. Based on the formulated aims,  Paas and Halapuu  \cite{TPaasVH012} used principal component factor analysis and micro-econometric methods data from the European Social Survey (ESS) fourth round database to examine the attitudes of European people towards immigrants.  
%The theories that explain the determinants of attitudes towards immigration are diverse. Generally, these theories can be divided into two groups: individual and collective theories. What distinguishes the two groups is the level of measurement; for example, country/region vs. person. Paas and Halapuu relied mainly upon individual economic theories (micro-approach) in considering the empirical focus. Based on interdisciplinary theoretical models for determinants of attitudes towards immigrants, they build a set of explanatory variables for estimated regression models for  attitudes towards immigrants taking into account country specific conditions \cite {TPaasVH012}. 
%The results of the empirical analysis are seen to be consistent with several theories explaining individual and collective determinants of people's attitudes towards immigrants. 
These attitudes of the European people's towards immigration -which strongly constraints mobility between regions- , these are proved to vary depending on several factors such as the personal characteristics of the respondents, the attitudes towards the country's institutions and socio-economic security, and, finally, country specific conditions. 
\cite{ch3}.  %N. K. Vitanov, . Ausloos, M.,  Knowledge epidemics and population dynamics models for describing idea diffusion, in {\it Models of Science Dynamics : Encounters Between Complexity Theory and Information Sciences}, Andrea Scharnhorst, Katy B\" orner, and Peter van den Besselaar, Eds. Springer Verlag Berlin Heidelberg  (2012) Ch. 3, pp. 69 -  125.

 On the analytic side, the work of Vitanov and Ausloos \cite{ch3} is noteworthy. Even though the authors are not considering a network \textit{per se}, they nevertheless include spatial gradients between regions in order to study issues such as: knowledge epidemics that take into account population dynamics and models describing the diffusion of ideas. This work relies on the use of the Lotka-Volterra system of equations with spatial gradients between regions with the addition of demographic input.
 
\section{Other social network models}
%\colorbox{yellow}{Please Marcel, remember to introduce the intro to sociophysics writing something on Galam and others}
%Traces of the application of quantitative methods to social sciences can be found in past history, but only  in more recent times the approach has been widely accepted and acknowledged. Approaches proposed be physicists, applied mathematicians, economists, social scientist, differ much each from the other due to the targets and methods of analysis. For instance, it is the law of big numbers which allows the application of statistical physics methods. On what concerns the application of networks for social modelling, they contribute to develop tools that allow social scientists to understand how and when social factors such as peer influences, role models, or norms affect individual choices.

%Network theory is increasingly gaining acceptance in the Economics community in order to understand the Economy as an evolving complex structure of interacting heterogeneous agents, the alternative view to that of the Economy as a homogeneous archipelago of independent \emph{homo economicus} underlying the neclassical mainstream. However, in Sociology and other disciplines like Ecology, Computation, etc. this has been so for decades now. In this Section we review briefly some interesting applications of complex networks to the description of social networks.
As outlined here above, network theory is increasingly gaining acceptance in the economics community in order to understand the Economy as an evolving complex structure of widely interacting heterogeneous agents  \cite{Ugomerloneetalchapter}.%(\colorbox{yellow}{chapter by Merlone et al.}). 
The alternative view to that is to consider the economy as an archipelago of  \emph{homo economicus} individuals, still interacting, but on a "shorter range", %yet rationally  leading to (structurally dissipative) patterns, thus 
underlying the neoclassical  economy \cite{Walras} mainstream features. However, in Sociology and other disciplines like Ecology, Computation, etc. this has been so for decades. 

In fact, the origin of considering a society as made of interacting entities goes back to Comte \cite{comte1852cours,comte1995leccons}, the founder of the discipline of sociology, having introduced the term as a neologism\footnote{The word was first coined by Sieys  in 1780 \cite{Guilhaumou}. }  in 1838, among other  scientific contributions.  Note that Comte had earlier used the term "social physics", but that term had been appropriated by others, e.g.   Quetelet \cite{quetelet1835homme,quetelet1869physique}. Thereafter,  Boltzmann and Maxwell imagined similar concepts for describing matter and natural phenomena.  Needless to say that much work has followed since.

Many examples of applications of the network approach to social sciences can be found in the literature. Reviews of this work can be found in Kirman \cite{kirman}, Stauffer \cite{stauffer2012biased}, in a compendium \cite{chakrabarti2007econophysics} and in the recent book \cite{galambook} further elaborating the work in Galam \cite{galam08}. Other studies have focused on computational techniques \cite{stauffer2003MCarlo}. In more recent times the approach has been widely accepted and acknowledged \cite{savoiu2012sociophysics}. The various methods used by physicists, applied mathematicians, economists, social scientists, differ much from each other due to the targets and methods of analysis. For instance, it is the law of large numbers which allows the application of statistical physics methods  \cite{stauffer2003MCarlo}. As far as the application of networks for social modelling is concerned, they contribute to develop tools that allow social scientists to understand how and when social factors such as peer influences, role models, or norms affect individual choices\cite{sousalmalarzgalam05,bischi2010global}. In what follows, several applications of complex networks are reviewed all aiming at describing some type of of social networks.

A large number of studies have applied complex networks to the study of systems like the Internet and the World Wide Web (WWW), and they have been extensively summarized in the reviews cited earlier. However, for an updated review focused specifically on applications the reader is referred to \cite{daFontouraCosta}, where applications of complex networks to real-world problems and data are reported. The authors surveyed the applications of complex networks formalism in "no less than 11 areas, providing a clear indication of the impact of the field of complex networks". Moreover, the book by Vega-Redondo \cite{Vega_redondo_book} provides a comprehensive coverage with applications of complex networks to labor markets, peer group effects, trust and trade, and research and development. 
%\bibitem Luciano da Fontoura Costa, Osvaldo N. Oliveira Jr., Gonzalo Travieso, Francisco Aparecido Rodrigues, Paulino Ribeiro Villas Boas, Lucas Antiqueira, Matheus Palhares Viana & Luis Enrique Correa Rocha Analyzing and modeling real-world phenomena with complex networks: a survey of applications, Advances in Physics 60, 3, 2011
%\bibitem{Vega_redondo_book} F. Vega-Redondo, Complex Social Networks. Cambridge University Press (January 8, 2007)
 
Social networks are known to be organized into densely connected communities, with a high degree of the clustering, and being highly assortative. The formation of complex networks has been reported form the experimental perspective by Bernasconi et al.\cite{Bernasconi} using non-cooperative games of network formation based of the Bala and Goyal type \cite{BalaGoyal}. Toivonen et al. \cite{Toivonen} reported a realistic model for an undirected growing network for its use in sociodynamic phenomena. On the other hand, Bogu\~n\'a et al. \cite{BogunaPastor} used an abstraction of the concept of social distance to define a class of models of social network formation. The evolution of structure within large online social networks is examined in \cite{KumarNovak}, with specific attention focused on the Flickr and Yahoo's social network, showing their segmentation, and providing a detailed characterization of the structure and evolution of their different regions as well as a simple model of network growth capable of mimicking this structure \cite{KumarNovak}. Finally, Palla et al. \cite{Pallanature} quantified social group evolution by means of an algorithm based on clique percolation for the time dependence of overlapping communities on a large scale. Finally, it is noteworthy that the problem of the determination of the community structure in the presence of  unobserved structures among the nodes -a rather common situation in social and economic networks- has been adressed by Copic et al. \cite{Kirman2}, who axiomatically introduced a maximum-likelihood-based method of detecting the latent community structures from network data.

Opinion formation in social systems has also been a matter of great concern in network literature. Apart from some pioneering work like that of Kirman and collaborators in the 80's using the diameter-2 Bollob\'as model \cite{Kirman1}, the field has evolved only recently when a plethora of contributions have been reported. As a very recent example, Koulouris et al. \cite{Koulouris} reported the multi-equilibria regulation of opinion formation dynamics.

In opinion formation models the single vote of an individual can be influenced and can change, but when the final target is the aggregate, the sampling of  many randomly selected people can give a reasonable impression for an upcoming election \cite{stauffer}. In economics agent-based models have been used for the analysis of the aggregate behaviour of a large number of individuals as model with heterogeneous agents are gaining a more prominent role relative to those with a representative agent \cite{kirman}.

Network theory has been also applied to other social networks of interest like opinion formation, social entrepreneurship, etc. Dal Forno and Merlone adapted the notion of density of a graph to multiple projects and non-dichotomous networks.  An appropriate visualization procedure has been implemented in \cite{3DalFornoMerlone08}.   
Social entrepreneurship effects on the emergence of cooperation in networks have been examined in \cite{4DalFornoMerlone08ECO11}, where differences between social entrepreneurs and leaders are analyzed and where the network of interactions may allow for the emergence of cooperative projects. The model reported by the authors consists on two coupled networks standing for knowledge and cooperation among individuals respectively. Any member of the community can be a social entrepreneur. On the basis of this theoretical framework, the authors prove that a moderate level of social entrepreneurship is enough for providing a certain coordination on larger projects, suggesting that a moderate level of social entrepreneurship would be sufficient. %While social leaders outperform social entrepreneurs in terms of number of links, their actions on the knowledge matrix are retained at a different level.
 
 %At this stage, the rest of this section contains only the raw abstracts of the contributions. It will be extended in following versions: 

Lambiotte and Ausloos in \cite{1JSM} analyzed the coexistence of opposite opinions in a network with communities. Applying the majority rule to a topology with two coupled random networks, they reproduced the modular structure observed in social networks.  The authors analytically calculated the asymptotic behavior of the model deriving a phase diagram that depends on the frequency of random opinion flips and on the inter-connectivity between the two communities. Three regimes were shown to take place: a disordered regime, where no collective phenomena takes place; a symmetric regime, where the nodes in both communities reach the same average opinion; and an asymmetric regime, where the nodes in each community reach an opposite average opinion, registering discontinuous transitions from the asymmetric regime to the symmetric regime. 

In this same model, Lambiotte et al. \cite{RLMAJH}  have shown that a transition takes place at a value of the interconnectivity parameter, above which only symmetric solutions prevail. Thus, both communities agree with each other and reach consensus. Below this value, the communities can reach opposite opinions resulting in an asymmetric state. They explicitly analyzed the importance of the interface between the subnetworks.

Finally Lambiotte and Ausloos \cite{RLMA494LCNS3993} studied collaborative tagging as a tripartite network, analyzing online collaborative communities  described by \emph{tripartite} networks whose nodes are persons, items and tags. Using projection methods they uncovered several structures of the networks, from communities of users to genre families. 

Finally, two economico-sociological studies are outlined. One pertains to the interaction of small world networks of biased communities, like the neocreationists $vs.$ the evolution defenders. For this analysis, the networks are considered to be directed but with unweighted links  \cite{garcia}   and   \cite{physaGR}. The other study  mentioned above pertains to bipartite networks made of music listeners downloading some music work  from the web \cite{music1,music2}.
 
\section{Suggestions for future research}

The application of the theory of complex networks, alone or in combination with other theoretical developments of statistical mechanics, can lead to very interesting results in several areas of economics in particular, and of sociophysics in general. Specifically, the elaboration of a model for the fluxes of entrepreneurs, trade and workers between the European regions undoubtedly demands the usage of complex networks together with non linear model for the dynamics of the agents. This treatment should generalize early regional science contributions, which are typically based on flow equations theories of directed diffusion (see for e.g. those in Refs. \cite{Hotelling,Beckman,Raa,Puu} %\colorbox{yellow}{must ask Tonu} 
where goods and migration fluxes are governed by conventional systems of diffusion-like partial differential equations). Moreover, this treatment could be very well complemented by a nonlinear model of production and consumption cycles, in the line of Meadows Dynamics of Commodity Production Cycles \cite{Meadows}. This could be done in analogy to what has been done for biological oscillators (see for example Goodwin's model of enzyme production in Refs. \cite{Goodwin,Murray} for a classical review on these kind of models). The introduction of spatial inhomogeneities in these models could also provide a new research path for economic geography. 

A list of other potentially fruitful research avenues is provided below.

{\bf Innovation and renewal of technology.}
It could be useful to apply complex networks and agent-based models to the analysis of the spreading of technological renewal, R\&D incentives and growth, fiscal  (regional) rules \cite{alison} usefulness of data analyses through rescale range analysis methods, principal component analysis. Moreover, this framework can also be applied to transportation, migration, growing and diversifying nodes of networks;  merging and controls of agents; or tendency toward monopoly through lobbying.

{\bf Regional trade and development.}
%{\em put a sentence here} text here
The network approach can also be applied for introducing relationships like trade barriers,community detection, clusters,  hierarchies; 
policy implications concerning the economic  (regional) clusters arising in the presence of Marshallian  and  other externalities, etc.

{\bf Development of database and data mining.}
{\em ``I would not have thought that the spread (IT/DE) was going to rise again'' [IT politician, summer 2012].}
Economic and financial theories need to be tested on real world markets. The complexity and large amount of data makes impossible
autonomous data collection at the individual level. Moreover, data providers as the Bureau van Dijk or Bloomberg do implement only certain type of data retrieval, and the work that researchers have to do autonomously delays the production of results, and the detection of information that can be used as input for more complex models. Moreover, the increasing costs of subscriptions
to data providers, in conjunction with the progressive decrease of national funding, suggest that the development of synergies with data providers is timely.

{\bf Statistical mechanics approaches}
In future works, phase transitions, coupling between magnetic and cristallographic transitions,  thermodynamic  (through the notion of cost function) vs. geometric (percolation) could be analyzed, including
% separation of space and time scales (necessity of averaging in  … optimal … time windows);
% external field effects (constraints);
%notion of endogeneous and exo-geneous effects
instabilities in necessarily non equilibrium structures (log-periodic oscillations). Moreover,
%nonlinear feedbacks     and avalanches
going beyond Ising model (Blume-Emery-Griffiths, and the forgotten ferroelectric models) should be considered, as well as community detection, forward and backward correlations in networks with weighted and directed links (danger of  difficulty in interpreting complex eigenvalues of adjacency matrix), network structure construction and evolution, etc.

{\bf Economic and financial networks and risk.}
{\em `` When Belgium sneezes, the world catches a cold'' []http://phys.org/news/2010-11-belgium-world-cold.html] }
Globalization of economic and financial markets, corporate ownership networks, international trade networks, as well as phenomena like tunnelling, cross-ownership, and boards interlock, change dramatically the profile of financial and economic risks pointing out the relevance of the network structure and are topics to be considered in the future. Understanding such phenomena both at the micro and macro level may help the development of policies also at the local level with potential benefits for regional trade and development.

\section{Conclusions}

The range of applications of complex networks formalism is expanding at a fabulous rate, and has been adopted almost in every field of knowledge having to deal with heterogeneous interacting agents and their emergent phenomena. This is the case, particularly, of economics, and even more specifically of economic geography. The present report gathers contributions from different fields and approaches under the common theme of complex networks analysis in socioeconomic models. The statistical mechanics of complex networks have been reviewed together with some computational aspects related to their description. Models specifically developed for examining topics in various areas of economics and finance, such as, for example, regional trade and development have been the object of specific attention, together with contributions devoted to the application of complex networks analysis to social networks in the broad sense.

\begin{flushleft}
{\bf Acknowledgment:} 
\end{flushleft}
This work has been performed in the framework of COST Action IS1104 
"The EU in the new economic complex geography: models, tools and policy evaluation".


\newpage
 \begin{thebibliography}{99} 
 
%  \bibitem{ER} P. Erd\"os and A. R\'enyi, The Evolution of Random Graphs, Magyar Tud. Akad. Mat. Kutat\'o Int. K\"ozl. 5 (1960) 17--61.

% \bibitem{BA} R. Albert, A.-L. Barab\'asi, Statistical Mechanics of Complex Networks, Rev. Mod. Phys. 74  (2002) 47--97.
 
\bibitem{euler}
Euler, L., (1736). Solutio problematis ad geometriam situs pertinentis. {\em Commentarii Academiae Scientiarum Imperialis Petropolitanae}, 8, 128-140.

\bibitem{Cayley} Cayley, J. (1889). A Theorem on Trees. \emph{Quarterly Journal of Mathematics} 23, 376–378

\bibitem{Erdos-Renyi} Erd\" os, P. \&  R\' enyi, A. (1959). On Random Graphs I. {\em Publications Mathematicae} 6,  290-297;  On the evolution of random graphs. \emph{Publications of the Mathematical Institute of the Hungarian Academy of Sciences (Magyar Tud. Akad. Mat. Kutat\'o Int. K\"ozl.)} 5, 17-61.

\bibitem{SWN} Watts, D.J. \&  Strogatz, S.H.  (1998). Collective dynamics of small-world networks. \emph{Nature}, 393, 440-442.

\bibitem{AB1999} Barab\'asi, A.-L. \&  Albert, R. (1999). Emergence of scaling in random networks. \emph{Science}, 286, 509-512.

\bibitem{reviewAB}  Albert R., \&  Barabasi, A.-L. (2002). Statistical Mechanics of Complex Networks. \emph{Review of Modern Physics}, 74, 47-97. 

\bibitem{newman0303516} Newman, M. E. J. (2003). The structure and function of complex networks. \emph{SIAM Reviews}, 45, 167-256. 

\bibitem{PastorSatorrasRDG} Pastor-Satorras, R., Rubi, M., D\'{\i}az-Guilera,  A. (Eds.) (2003). \emph{Statistical Mechanics of Complex Networks}. Berlin: Springer.

\bibitem{reviewBocaletti} Boccaletti ,S., Latora, V., Moreno, Y., Chavez, M., Hwang, D.-U. (2006). Complex networks: Structure and dynamics. \emph{Physics Reports}, 424, 175-308.

\bibitem{reviewnewmanwatts} Newman, M.,Watts, D., Barab\'asi, A.-L. (2006) \emph{The Structure and Dynamics of Networks}. Princeton (New Jersey): Princeton University Press. 

\bibitem{496RLMAwcss} Lambiotte R., \& Ausloos, M. (2006). Modeling the evolution of coupled networks, \emph{First World congress on Social Simulation. e-Proceedings}, 1, 375-381.

\bibitem{520MARLdrastic}   Ausloos, M. \&  Lambiotte, R (2007). Drastic events make evolving networks. \emph{Eureropean Physics Journal B}, 57, 89-94.

\bibitem{511RLMAj}  Lambiotte, R.  \& Ausloos, M. (2007) Growing network with j-redirection, \emph{Europhyics Letters}, 77, 58002. %$ %(http://arxiv.org/abs/physics/0612148, 35kb) $

\bibitem{Ugomerloneetalchapter} M. Ausloos,  H. Dawid, \& U. Merlone, (2014). Spatial interactions in ABM. {\it a chapter in this book}.

\bibitem{NamatameKA} Namatame, A., Kaizouji, T.,  Aruka, Y. (Eds.) (2006). \emph{The complex networks of economic Interactions}.  Berlin: Springer.

\bibitem{AndersonArrowPines} %Philip W. Anderson, Kenneth Arrow, David Pines, Eds., The Economy as an evolving complex system, vols I-II (Westview Press, \colorbox{green}{lugar} 1988)

%======= or =======

Barab\' asi, A.-L. (2003) \emph{Linked How everything is connected to everything else and what it means  for Business, Science, and Everyday Life}, Plume Books.

\bibitem{BlumeDurlauf} %L.E. Blume, S. Durlauf Eds., The Economy as an evolving complex system, vol. III (Oxford Universitty Press, 2006)
%======= or ========
Dorogovtsev, S. N. \& Mendes, J. F. F. (2003) \emph{Evolution of Networks - From
Biological Nets to the Internet and WWW.} Oxford: Oxford University Press.


%\bibitem{Action} In this chapter, special attention is paid to contributions of the members of COST Action IS1104 "The EU in the new economic complex geography: models, tools and policy evaluation".

\bibitem{Avrutin} Avrutin, V., Levi, P., Schanz, M., Fundinger, D., Osipenko, G.S. (2006). Investigation of dynamical systems using symbolic images: efficient implementation and applications. \emph{International Journal of Bifurcation and Chaos}, 16, 3451-3496.

\bibitem{symbolic}Rotundo, G. (2012) An investigation of computational complexity of the method of symbolic images. In: \emph{Advanced Dynamic Modelling of Economic and Social Systems. Studies in computational intelligence}, A. Proto, M. Squillante, J. Kacprzyk (Eds.) (pp. 109-126). Berlin: Springer.

\bibitem{daCostareview} Costa, L. da F., Rodrigues, F. A., Travieso, G., Villas Boas, P. R. (2007) Characterization of complex networks: A survey of measurements. \emph{Advances in Physics}, 56, 167–242.

\bibitem{hidden} Serrano, M. A., Krioukov, D., Bogu\~n\'a, M. (2008). Self-Similarity of Complex Networks and Hidden Metric Spaces. \emph{Physical Review Letters}, 100, 078701. 

\bibitem{Milgram1967}  Milgram, S. (1967). The small-world problem.\emph{ Psychology Today}, 1, 60-67.

\bibitem{Djikstra} Dijkstra, E. W. (1959). A note on two problems in connexion with graphs. \emph{ Numerische Mathematik}, 1, 269–271.

\bibitem{gligorausloosAOI} Gligor, M. \& Ausloos, M. (2008).
 Clusters in weighted macroeconomic networks: the EU case. Introducing the overlapping index of  GDP/capita fluctuation correlations. \emph{European Physical Journal B}, 63,  533-539.
 
\bibitem{betweenness} Freeman, L. (1977). A set of measures of centrality based on betweenness. \emph{Sociometry}, 40, 35-41.

\bibitem{correlations}  Barrat, A., Barthélemy, M., Pastor-Satorras, R., Vespignani, A. (2004). The architecture of complex weighted networks. \emph{Proceedings of the National Academy of Sciences}, 101, 3747-3752.
 
\bibitem{assortativity} Newman, M. E. J. (2002). Assortative Mixing in Networks.
\emph{Physical review Letters}, 89, 208701.
%\bibitem{mixing} M.E.J. Newman,  {\it Mixing patterns in Networks}, arXiv:cond-mat/0209450v2 Phys. Rev. E {\bf 67}   (2003)  026126. %[13 pages]

\bibitem{Boguna}  Bogu\~n\'a, M. \& Pastor-Satorras, R. (2002). Epidemic spreading in correlated complex networks. \emph{Physical Review E}, 66, 047104.

 \bibitem{Panos} Garas, A., Argyrakis, P., Rozenblat, C., Tomassini, M., Havlin, S. (2010) Worldwide spreading of economic crisis. \emph{New Journal of Physics}, 12, 113043.
 
 \bibitem{Garas} Garas, A., Schweitzer, F., Havlin, S. (2012). A k-shell decomposition method for weighted networks. \emph{New Journal of Physics}, 14, 083030.

 \bibitem{AraujoPhA389} Araujo, A.I.L., Corso, G., Almeida, A.M., Lewinsohn, T.M. (2010). An analytic approach to the measurement of nestedness in bipartite networks. \emph{Physica A}, 389, 1405-1411.

\bibitem{Almeida-Neto}  Almeida-Neto, M., Guimar\~{a}es, P., Guimar\~{a}es, Jr., P.R., Loyola, R.D., Ulrich, W. (2008). A consistent metric for nestedness analysis in ecological systems: reconciling concept and measurement.\emph{Oikos}, 117, 1227-1239.

\bibitem{NewmanMooreWatts} Newman, M.E.J., Moore, C., Watts, D.J. (2000). Mean-field solution of the small-world network model. \emph{Physical Review Letters}, 84, 3201-3204.

\bibitem{Ising} Ising, E. (1925).Beitrag zur Theorie des Ferromagnetismus,\emph{ Zeitschrift f\"ur Physik}, 31, 253–258.

\bibitem{Gitterman} Gitterman, M. (2000). Small-world phenomena in Physics: the Ising model. \emph{Journal of Physics A: Mathematical and General}, 33, 8373-8381.

\bibitem{Barrat}  Barrat, A., Weigt, M. (2000). On the properties of small-world network models. \emph{European Physics Journal B}, 13, 547-560.

\bibitem{Pekalski}  Pekalski, A. (2001). Ising model on a small world network.
\emph{Physical Review E}, 64,057104.

\bibitem{Viana-Lopes} Viana-Lopes, J., Pogorelov, G., Lopes dos Santos, J.M.B, Toral, R. (2004) Exact solution of Ising model on a small-world network. \emph{Physical Review E}, 70, 026112.

\bibitem{Herrero} Herrero, C.P. (2002). Ising model in small-world networks, \emph{Physical Review E}, 65, 066110.

\bibitem{Bianconi} Bianconi, G. (2002). Mean field solution of the Ising model on a Barab\'asi-Albert network. \emph{Physics Letters A}, 303, 166-168.

\bibitem{Onsager1944} Onsager, L. (1944). Crystal statistics. I. A Two-Dimensional model with an order-disorder transition, \emph{Physical Review}, 65, 117–149.

\bibitem{LeBellac} LeBellac, M. (1992). \emph{Quantum and statistical field theory}. New York: Oxford University Press.

\bibitem{Yang} Yang, C. N. (1952). The spontaneous magnetization of a two dimensional Ising model. \emph{Physical Review}, 85, 808-816.

\bibitem{Mattis} Mattis, D. C.  (1976). Solvable spin systems with random interaction. \emph{Physics Letters }, 56A, 421-422.

 \bibitem{467RLMApre72}   Lambiotte, R., Ausloos, M. (2005)  $N$-body decomposition of bipartite networks, \emph{Physical Review E}, 72, 066117.

\bibitem{Hotelling} Hotelling, H. (1929). Stability in Competition. Economic Journal, 39, 41-57.

\bibitem{garlaschelli}Caldarelli, G.,  Battiston, S., Garlaschelli, D., Catanzaro, M. (2004). Emergence of Complexity in Financial Networks. \emph{Lecture Notes in Physics: Complex Networks}, 650, 399-423.
 
\bibitem{bertone} Bertoni, F. \&  Randone, P. A. (2006). The Small-World of Italian Finance: Ownership Interconnections and Board Interlocks Amongst Italian Listed Companies, \emph{Technical Report, Politecnico di Milano}.

\bibitem{GRAMDAASS}  Rotundo, G., D'Arcangelis, A.M. (2010). Network analysis of ownership and control structure in the Italian Stock market. \emph{Advances and applications in statistical sciences}, 2, 255-273.

\bibitem{grassi}  Grassi, R. (2010). Vertex centrality as a measure of information flow in Italian Corporate Board Networks, \emph{Physica A}, 389, 2455-2464.

 \bibitem{croci} Croci, E. \& Grassi, R. (2013)  The economic effect of interlocking directorates in Italy: new evidence using centrality measures. \emph{Computational and Mathematical Organization Theory}, in press.

\bibitem{music1} Lambiotte, R. \& Ausloos, M. (2005). Uncovering collective listening habits and music genres in bipartite networks, \emph{Physical Review E}, 72, 066107.

\bibitem{music2} Lambiotte, R., Ausloos, M. (20006) On the genrefication of Music: a percolation approach. \emph{European Physics Journal B}, 50, 183-188.

\bibitem {CPC147.02.40Newman}  Newman, M.E.J. (2002). The structure and function of networks. \emph{Computer Physics Communications}, 147, 40-45. 
 
 \bibitem{BoostGraph}  Siek, J.G., Lee, L.-Q., Lumsdaine, A. (2001). \emph{The Boost Graph Library.} Reading (Massachusetts): Addison-Wesley.

 \bibitem{sparse} Pissanetzky, S. (1984). \emph{Sparse Matrix Technology}. New York:Academic Press.

 \bibitem{Gilbert}  Gilbert, E.N. (1959). Random Graphs. \emph{Annals of  Mathematical Statistics}, 4, 1141-1144.

 \bibitem{Facebook}  Ugander, J.,  Karrer, B.,  Backstrom, L., Marlow, C. (2011). The Anatomy of the Facebook Social Graph, \emph{CoRR} abs/1111.4503.

 \bibitem{genBA} Krapivsky, P.L., Redner, S., Leyvraz, F. (2000) Connectivity of growing random networks. \emph{Physical Review Letters}, 85, 4629-4632.

 \bibitem{merger} Seyed-allaei, H., Bianconi, G., Marsili, M. (2006). Scale-free Networks with an Exponent less than Two. \emph{Physical Review E}, 73, 046113.

 \bibitem{exponentialorscalefree} Cotilla-Sanchez, E., Hines, P.D.H., Barrows, C., Blumsack, S. (2012) Comparing the Topological and Electrical Structure of the North American Electric Power Infrastructure. \emph{IEEE Systems Journal }, 6, 616-626.

 \bibitem{generalized}  Newman, M.E.J., Strogatz S.H., Watts, D.J. (2001). Random Graphs with Arbitrary Degree Distributions and their Applications. \emph{Physical Review E}, 64, 026118.

 \bibitem{molloy1} Molloy, M., Reed,  B. (1995) A critical point for random graphs with a given degree sequence. \emph{Random Structures and Algorithms}, 6, 161-179.

 \bibitem{molloy2} Molloy, M., Reed, B. (1998) The size of the giant component of a random graph with a given degree sequence. \emph{Combinatorics, Probability and Computing} 7, 295-305.

  \bibitem{pajek} Batagelj, V., Mrva, A. (2003). Pajek$-$Analysis and Visualization of Large Networks, in  M. J\"unger and P. Mutzel (Eds.) (pp. 77-103). Berlin:Springer.
  
  \bibitem{gephi} Bastian, M.,  Heymann,  S., Jacomy, M. (2009). Gephi: an open source software for exploring and manipulating networks, \emph{International AAAI Conference on Weblogs and Social Media}.

 \bibitem{python} \emph{The Python Tutorial}, http://docs.python.org/2/tutorial.  
 
 \bibitem{Tesfatsion} Tesfatsion, L. (2003). Agent-based computational economics: modelling economies as complex adaptive systems. \emph{Information Sciences}, 149, 262-268.
 
\bibitem{Kirman1997} Kirman, A. (1997) The economy as an evolving network.  \emph{Journal of Evolutionary Economics}, 7, 339-353.

\bibitem{BarkleyRosser1999} Barkley Rosser Jr. J. (1999). On the Complexities of Complex Economic Dynamics. \emph{The Journal of Economic Perspectives} 13, 169-192.

\bibitem{ColanderHoltRosser} Colander, D., Holt, R., Barkley Rosser Jr., J. (2004).  The changing face of mainstream economics.\emph{Review of Political Economy}, 16, 485-499.

\bibitem{BrianArthur2006}  Brian Arthur, W. (2006). Out-of-Equilibrium Economics and Agent-Based Modelling. \emph{Handbook of Computational Economics}, 2, 1551–1564.

\bibitem{Foster2005} Foster, J. (2005). From simplistic to complex systems in economics. \emph{Cambridge Economics Journal}, 29, 873-892.

\bibitem{MartinSunley} Martin, R. \& Sunley, P. (2007) Complexity thinking and evolutionary economic geography. \emph{The Journal of Economic Geography}, 7, 573-601.

\bibitem{MJackson} Jackson, M. O. (2011) An Overview of Social Networks and Economic Applications, in \emph{The Handbook of Social Economics}, J. Benhabib, A. Bisin, and M.O. Jackson (eds.), North Holland Press.

\bibitem{Bhattacharya} Bhattacharya, K., Mukherjee, G., Saramaki, J., Kaski, K. and Manna, S. S. (2008) The International Trade Network, in \emph{Econophysics of Markets and Business Networks, New Economic Windows Series}(pp. 139-147). Berlin:Springer.

\bibitem{Caldarellinature} Caldarelli, G., Chessa, A., Gabrielli, A., Pammolli, F., Puliga, M. (2013). Reconstructing a credit network. \emph{Nature Physics}, 9, 119-197.

\bibitem{daCruz} da Cruz, J.P. \&  Lind, P.G. (2012). The dynamics of financial stability in complex networks. \emph{European Physics Journal B}, 85, 256-265.

\bibitem{Kali} Kali, R., Reyes, J. (2007) The Architecture of Globalization: A Network Approach to International Economic Integration. \emph{Journal of International Business Studies}, 28, 595-620.

\bibitem{Rodrigue} Rodrigue, J.-P. (2013). \emph{Transportation, Globalization and International Trade}. New York: Routledge. 

\bibitem{Xiang_Lia} Xiang Lia, Yu Ying Jin, Guanrong Chen (2003). Complexity and synchronization of the World trade Web. \emph{Physica A}, 328, 287–296.

\bibitem{Garlaschelli2} Garlaschelli, D., Loffredo, M. I. (2005). Structure and evolution of the world trade network. \emph{Physica A}, 355, 138–144.

\bibitem{Fagiolo}  Schweitzer, F., Fagiolo,  G., Sornette, D., Vega-Redondo,F., White, D. R. (2009). Economic Networks: What do we know and what do we need to know? \emph{Advances in Complex Systems}, 12, 4.

\bibitem{FujiwaraAoyama} Fujiwara, Y. \&  Aoyama, H. (2010). Large-scale structure of a nation-wide production network. \emph{The European Physical Journal B}, 77(4), 565-580. 

\bibitem{LeeK-M} Lee K-M, Yang J-S, Kim G, Lee J, Goh K-I, et al. (2011). Impact of the Topology of Global Macroeconomic Network on the Spreading of Economic Crises. \emph{PLoS ONE} 6(3): e18443. doi:10.1371/journal.pone.0018443.

\bibitem{475NetMisk}  Miskiewicz, J. \&  Ausloos, M. (2006). G7 country Gross Domestic Product (GDP) time correlations. A graph network analysis,  in \emph{Practical Fruits of Econophysics}, H. Takayasu, (Ed.) (pp. 312-316) Tokyo: Springer. %$  (http://arxiv.org/abs/physics/0504099,  13kb)  $

\bibitem{512MGMAepjb57} Gligor, M. \& Ausloos, M. (2007) Cluster structure of EU-15 countries derived from the correlation matrix  analysis of macroeconomic index fluctuations.\emph{ European Physical Journal B}, 57, 139- 146.  

\bibitem{514MARLPhA382} Ausloos,  M.  \&  Lambiotte, R. (2007). Clusters or networks of economies? A macroeconomy study through GDP fluctuation correlations. \emph{Physica A}, 382, 16-21.
%$ (http://arxiv.org/pdf/physics/0607139, 266kb) $
 %doi:10.1016/j.physa.2007.02.005   
 
\bibitem{537} Gligor, M. \&  Ausloos, M. (2008). Convergence and cluster structures in EU area according to fluctuations in macroeconomic indices. \emph{Journal of Economic Integration}, 23, 297-330.  

\bibitem{539}  Gligor, M. \&  Ausloos, M. (2008). Clusters in weighted macroeconomic networks : the EU case. Introducing the overlapping index of GDP/capita fluctuation correlations. \emph{ European Physical Journal B}, 63, 533-539.

\bibitem{544}  Gligor, M. \&  Ausloos, M. (2008), Cluster Expansion Method for Evolving Weighted Networks Having Vector-like Nodes.  \emph{Acta Physica Polonica A},114, 491-499.

\bibitem{551}   Redelico,  F. O., Proto,  A.N., Ausloos, M. (2009). Hierarchical structures in the Gross Domestic Product per capita fluctuation in Latin American countries. \emph{Physica A}, 388, 3527-3535.

\bibitem{BougheasKirmanChapter}   Bougheas, S. \&  Kirman, A. (2014).  Complex financial networks and systemic risk: a review, {\it a chapter in this book}.
 
 \bibitem{options} Cerqueti, R. \& Rotundo, G (2010). Options with underlying asset driven by a fractional brownian motion: crossing barriers estimates. \emph{New Mathematics and Natural Computation}, 6, 109-118.
 
 \bibitem{levy} Levy, H., Levy, M., Solomon, S. (2000) \emph{Microscopic Simulation of Financial Markets: From Investor Behavior to Market Phenomena}. Academic Press. 
 
  \bibitem{Mantegna1999} Mantegna, R.N. (1999). Hierarchical structure in financial markets. \emph{European Physical Journal B}, 11, 193-197. %http://dx.doi.org/10.1007/s100510050929.  http://arxiv.org/pdf/cond-mat/9802256.pdf
  
 \bibitem{pozzi} Pozzi, F., Aste, T. Rotundo, G., Di Matteo, T. (2008). Dynamical correlations in financial systems. In: \emph{Complex Systems II. Proceedings of the SPIE, The International Society for Optical Engineering}, 6802, 68021E.
 
 \bibitem{calda} Caldarelli, G., Lillo, F., Mantegna, R.N. (2003). Topology of correlation-based minimal spanning trees in real and model markets.  \emph{Physical Review E}, 68, 046130.
  
\bibitem{pozzi2}
Pozzi, F., Di Matteo, T., Aste, T. (2013). Spread of risk across financial markets: better to invest in the peripheries. \emph{Scientific Reports} 3, 1665.

\bibitem{GRAMDQQ} Rotundo, G.,  D'Arcangelis, A.M., Network of firms: an analysis of market concentration. \emph{Quality \& Quantity}, in press.

 \bibitem{GRAMDJEIC} Rotundo, G.,  D'Arcangelis, A.M. (2010). Ownership and control in shareholding networks. \emph{Journal of Economics Interaction and Coordination}, 5, 191-219. 

 \bibitem{DErrico} D'Errico M., Grassi R., Stefani S., Torriero A. (2008). Shareholding Networks and Centrality: an Application to the Italian Financial Market, in A. Naimzada, S. Stefani, A. Torriero (Eds.) \emph{Network, Topology and Dynamics. Theory and Applications to Economics and Social Systems} (pp. 215-228). Berlin:Springer.
 
 \bibitem{GRNATO} Rotundo, G. (2011). Centrality Measures in Shareholding Networks. In: \emph{Use of Risk Analysis in Computer-Aided Persuasion. NATO Science for Peace and Security Series}, 88, 12-28, Amsterdam:IOS Press.
 
 \bibitem{Rotundo} Rotundo, G., On centrality measures for shareholding networks: measuring the spreading power of crises, preprint.
 
 \bibitem{salvemini}
Salvemini, M.T., Simeone, B., Succi, R. (1995). Analisi del possesso integrato nei gruppi di imprese mediante grafi. \emph{L’Industria} XVI(4), 641-662.
 
%\bibitem{Panos} A. Garas, P. Argyrakis, C. Rozenblat, M. Tomassini, and S. Havlin, Worldwide spreading of economic crisis, New Journal of Physics 12 (2010) 113043.
 
% \bibitem{Garas} A. Garas, F. Schweitzer, S. Havlin (2012). A k-shell decomposition method for weighted networks, New Journal of Physics 14 (8) 083030.
 
\bibitem{Vitali} Vitali,  S.,  Glattfelder, J.B., Battiston, S. (2011). The Network of Global Corporate Control. \emph{PLoS ONE}, 6(10) e25995. 
 
\bibitem{LuxWesterhoff} Lux, T \& Westerhoff, F. (2009). Economic crisis. \emph{Nature Physics}, 5, 2-3. 

\bibitem{Catanzaro} Catanzaro, M., Buchanan, M. (2013). Network opportunity. \emph{Nature Physics}, 9, 121-122.

\bibitem{Galbiati} Galbiati, M., Delpini, D., Battiston, S. (2013). The power to control. \emph{ Nature Physics}, 9, 126-128.

\bibitem{Onnela1} Onnela, J.-P., Chakraborti, A., Kaski, K., Kert\'esz, J.  (2002). Dynamic asset trees and portfolio analysis. \emph{European Physical Journal B}, 3, 285.

\bibitem{Onnela2} Onnela, J.-P., Chakraborti, A., Kaski, K., Kert\'esz, J. (2003).  Dynamic asset trees and Black Monday. \emph{Physica A}, 324, 247.

\bibitem{Onnela3} Onnela, J.-P., Chakraborti, A., Kaski, K., Kert\'esz, J., Kanto, A. Asset trees and asset graphs in ﬁnancial markets. \emph{Physica Scripta}, 106, 48.

\bibitem{Onnela4} Onnela, J.-P.,Chakraborti, A., Kaski, K, Kert\'esz, J., Kanto, A. (2003). Dynamics of market correlations: Taxonomy and portfolio analysis. \emph{Physical Review E}, 68, 056110.

\bibitem{Onnela5} Onnela, J.-P., Kaski, K., Kert\'esz, J. (2004). Clustering and information in correlation based ﬁnancial networks. \emph{European Physical Journal B}, 38, 353.

\bibitem{Onnela6} Onnela, J.-P., J. Saramäki, Kert\'esz, J., Kaski, K. (2005). Intensity and coherence of motifs in weighted complex networks. \emph{Physical Review E}, 71, 065103.

\bibitem{Onnela7} J. Saramäki, Onnela, J.-P., Kert\'esz, J., Kaski, K. (2005). Characterizing Motifs in Weighted Complex Networks in: Science of Complex Networks,J.F.F. Mendes, et al. (Eds.) (p. 108). AIP Conference Proceedings.

\bibitem{Onnela8} Onnela, J.-P., Saramäki, J., Kaski, K., Kert\'esz, J. (2006). Financial market- a network perspective in: H. Takayasu (Ed.)\emph{ Practical Fruits of Econophysics, Nikkei Econophysics III Proceedings} (p. 302). Tokyo:Springer.

\bibitem{Onnela9}PhD Thesis Onnela, J.-P.. Complex networks in the study if financial and social systems. $http://jponnela.com/web$\_$documents/t2.pdf$

\bibitem{Oatley} Oatley, T., Winecoff, W.K., Pennock,  A., Danzman, S.B. (2013). The Political Economy of Global Finance: A Network Model. \emph{Perspectives on Politics} 1, 133-153.

\bibitem{JVA}  Vitting Andersen, J.,  Nowak, A. Rotundo, G., Parrott,  L., Mart\'inez, S. (2011). "Price-Quakes" Shaking the World’s Stock Exchanges. \emph{PLoS ONE} 6, e26472 .

\bibitem{Bonanno}  Bonanno, G., Caldarelli, G.,  Lillo, F., Micciché, S., Vandewalle, N., Mantegna, R. N. (2004). \emph{European Physical Journal B}, 38, 363-371.

\bibitem{Westerhoff08PhA387}
Zaklan, G., Lima, W., Westerhoff, F. (2008) Controlling tax evasion fluctuations. \emph{Physica A}, 387, 5857-5861.

\bibitem{ZWesterhoffS09JEIC4} Zaklan, G., Westerhoff, Stauffer,  F.D. (2009). Analysing tax evasion dynamics via the Ising model. \emph{Journal of Economics Interaction and Coordination}, 4, 1-14.

\bibitem{Dunia1} L\'opez-Pintado, D. (2008). Diffusion in Complex Social Networks. \emph{Games and Economic Behavior}, 62, 573-590.

\bibitem{Dunia2}  L\'opez-Pintado, D. (2008). The Spread of Free-Riding Behavior in a Social Network. \emph{Eastern Economic Journal}, 34, 464–479. 

\bibitem{Dunia3} L\'opez-Pintado, D. \& Watts, D. J. (2008). Social Influence, Binary Decisions and Collective Dynamics.\emph{ Rationality and Society}, 20, 399-443.

\bibitem{Semitiel} Semitiel-Garc\'ia, M. \& Noguera-M\'endez, P. (2012). The structure of inter-industry systems and the diffusion of innovations: The case of Spain. \emph{Technological Forecasting and Social Change}, 79, 1548-1567.

\bibitem{Souma} Souma, W., Fujiwara, Y., Aoyama, H. (2003). Growth and Fluctuations of Personal and Company's Income. \emph{Physica A}, 324, 396–401.

\bibitem{RotundoCerquetiPuu}
Rotundo, G., Cerqueti,  R. (2010) Firms clustering in presence of technological renewal processes. In: T. Puu, A. Panchuk. \emph{Nonlinear Economic Dynamics}, (pp. 135-140) :Nova Science Publishers.

\bibitem{CRITOR} Cerqueti, R., Rotundo, G. (2007). Productivity and costs for firms in presence of technology renewal processes, \emph{International Transactions in Operational Research}, 14, 521–534.

\bibitem{RotundoCerquetiAMS}
Cerqueti, R.,  Rotundo, G. (2009) Companies' Decisions for Profit Maximization: A Structural Model. \emph{Applied Mathematical Sciences}, 3, 1327-1340. 

\bibitem{amaral1} Amaral, L.A.N., Buldyrev, S.V., Havlin, S., Leschhorn, H., Maass, P., Salinger, M.A., Stanley, H.E., Stanley, M.H.R. (1997).  Scaling behavior in economics: I. Empirical results for company growth.\emph{Journal de Physique I}, 7, 621–633.

\bibitem{amaral2} Amaral, L.A.N., Buldyrev, S.V., Havlin, S., Leschhorn, H., Maass, P., Salinger, M.A., Stanley, H.E., Stanley, M.H.R. (1997). Scaling behavior in economics: II. Modeling of company growth. \emph{Journal de Physique I}, 7, 635–650.

\bibitem{axtell}Axtell, R. L. (2001). Zipf distribution of U.S. firm sizes. \emph{ Science}, 293, 1818–1820.

\bibitem{BischiLamanitaI12}
Bischi, G.I., Lamantia, F. (2012). A dynamic model of oligopoly with R\&D externalities along networks. Part I. \emph{Mathematics and Computers in Simulation}, 84, 51-65.

\bibitem{BischiLamanitaII12}
Bischi, G.I., Lamantia, F. (2012) A dynamic model of oligopoly with R\&D externalities along networks. Part II. \emph{Mathematics and Computers in Simulation}, 84, 66–82.

\bibitem{KManasakisZ12JEMS}
Kesavayuth, D., Manasakis, C., Zikos, V., Upstream R\&D Networks and Welfare, working paper %submitted to \emph{Journal of Economics and Management Strategy}.

\bibitem{ManasakisPZ12SJE}
Manasakis, C., Petrakis, E., Zikos, V., Downstream Research Joint Venture with upstream market power,  working paper %submitted to \emph{The Southern Journal of Economics}.

%\bibitem{Glattfelder} J. B. Glattfelder, S. Battiston, Backbone of complex networks of corporations: the flow of control, \emph{Physical Review E} 80 (2009) 036104 
 
 \bibitem{2DalFornoMerlone2007}
Dal Forno, A., Merlone, U.  (2007).  The evolution of coworkers networks: An experimental and computational approach, in Edmonds, B., Hern\'andez Iglesias, C., Troitzsch, K.G., (Eds.), \emph{Social Simulation: Technologies, Advances and New Discoveries, Information Science Reference} (pp. 280-293). Hershey (PA).

\bibitem{pombo} Pombo-Romero, J., Varela, L.M., Ricoy, C. (2013) Diffusion of innovations in social interaction systems. An agent-based model for the introduction of new drugs in markets. \emph{European Journal of Health Economics}, 14, 443-455.
 
\bibitem{Ausloos08}  Ausloos, M., Lambiotte, R., Scharnhorst,  A., Hellsten, I. (2008). Andrzej Pekalski networks of scientific interests with internal degrees of freedom through self-citation analysis. \emph{International Journal of Modern Physics C}, 19, 371-384.  
 
 \bibitem{Iina06}  Hellsten, I., Lambiotte, R., Scharnhorst, A., Ausloos, M. (2006). A journey through the landscape of physics and beyond - the self-citation patterns of Werner Ebeling. in:\emph{ Irreversible Prozesse und Selbstorganisation.} T.  Poeschel, H.  Malchow  and L. Schimansky-Geier (Eds.) (pp. 375-384). Berlin:Logos Verlag.

\bibitem{Iina07} Hellsten, I., Lambiotte, R., Scharnhorst, A., Ausloos, M. (2007).  Self-citations, co-authorships and keywords: A new method for detecting scientists' field mobility? \emph{Scientometrics},72, 469-486.

\bibitem {Fronczak} Fronczak, A. \& Fronczak, P. (2012). Statistical mechanics of the international trade network. \emph{Physical Review E}, 85, 056113.

\bibitem{Fagiolo_Reyes} Fagiolo, G., Reyes, J., Schiavo, S. (2008). On the topological properties of the world trade web: A weighted network analysis. \emph{Physica A}, 387, 3868–3873. 

\bibitem{Barigozzi} Barigozzi, M., Fagiolo, G., Mangioni, G. (2011). Community Structure in the Multi-network of International Trade Complex Networks. \emph{Communications in Computer and Information Science}, 116, 163-175. 

\bibitem{ReyesWooster} Reyes, J.A., Wooster, R. B., Shirrell, S. (2009). Regional Trade Agreements and the Pattern of Trade: A Networks Approach. Available at SSRN: http://ssrn.com/abstract=1408784 or http://dx.doi.org/10.2139/ssrn.1408784.

\bibitem{ReyesSchiavo} Reyes, J., Schiavo, S., Fagiolo, G. (2010). Using complex networks analysis to assess the evolution of international economic integration: The cases of East Asia and Latin America, \emph{The Journal of International Trade and Economic Development}, 19, 215–239.

\bibitem{PaasS08ijrs7} Paas, T. and Schlitte, F. (2008) Regional income inequality and convergence process in the EU-25. \emph{Scienze Regionali. Italian Journal of Regional Science}, 7, 29-49. 

\bibitem{TPaasTV012} Paas, T., Vahi, T.  Regional Disparities and Innovations in Europe,  http://ideas.repec.org/p/wiw/wiwrsa/ersa12p80.html

\bibitem{TPaasVH012} Paas, T., Halapuu, V. (2012). Attitudes towards immigrants and the integration of ethnically diverse societies, \emph{Norface Migration Discussion Paper} No. 2012-23. 

\bibitem{ch3} Vitanov, N. K., Ausloos, M. (2012). Knowledge epidemics and population dynamics models for describing idea diffusion, in {\it Models of Science Dynamics : Encounters Between Complexity Theory and Information Sciences}, Andrea Scharnhorst, Katy B\" orner, and Peter van den Besselaar, (Eds.) (pp. 69-125). Berlin:Springer.
 
 \bibitem{Walras}  Walras, L. (1954)  Elements of pure economics : or, The theory of social wealth. London:Allen \& Unwin.

 \bibitem{comte1852cours} Comte, A. (1852). Cour de philosophie positive. Paris:Borrani et Droz.
 
  \bibitem{comte1995leccons}  Comte, A. (1995)   Le\c cons sur la sociologie: Cour  de philosophie positive: le\c cons 47 \`a 51.  Paris:Juliette Grange Flammarion.
  
 \bibitem{Guilhaumou}Guilhaumou,  J. (2006) Siey\'es et le non-dit de la sociologie : du mot  la chose.  \emph{Revue d'histoire des sciences humaines,  Naissance de la science sociale (1750-1850)},  15, 117-134
 
  \bibitem{quetelet1835homme} Quetelet,  A. (1835).  Sur l'homme et le d\' eveloppement de ses facult\' es, ou Essai de physique sociale, Paris:Bachelier; English translation (1842),  A Treatise on Man and the Development of His Faculties  Edinburgh: William and Robert Chambers.
  
    \bibitem{quetelet1869physique}   Quetelet, A. (1869).  Physique sociale, ou essai sur le d{\'e}veloppement des facult{\'e}s de l'homme, Paris:Muquard;  Engl. transl. (Franklin,  1968).
    
  %\bibitem{kirmanJEP92} Kirman, A. (1992). Whom of What Does the Representative Individual Represent?. \emph{Journal of Economic Perspectives}, 6, 117-136.

    \bibitem{stauffer2012biased}
Stauffer, D. (2012). A Biased Review of Sociophysics. \emph{Journal of Statistical Physics}, 151, 9-20.

  \bibitem{chakrabarti2007econophysics} Chakrabarti, B.K., Chakraborti, A.,  Chatterjee, A.(2007) Econophysics and sociophysics. Wiley. 
  
     \bibitem{galambook}  Galam, S. (2012) \emph{What is Sociophysics About?}, Berlin:Springer.
 
  \bibitem{galam08}   
 Galam, S. (2008) Sociophysics: a review of Galam models. \emph{International Journal of Modern Physics C}, 19,409-440.
     
 \bibitem{stauffer2003MCarlo}
 Stauffer, D., (2003). Sociophysics. A review of recent Monte Carlo simulations. \emph{Fractals}, 11, 313-318. %Bonabeau et al. for the formation of social hierarchies, Donangelo and Sneppen for the replacement of barter by money, Solomon and Weisbuch for marketing percolation, and Sznajd for political persuasion. Finally we review how to destroy the internet.

 \bibitem{savoiu2012sociophysics} S\u{a}voiu, Gh., Iorga-Sim\u{a}n, I. (2012).
Sociophysics: A New Science or a New Domain for Physicists in a Modern University, in \emph{Econophysics: Background and Applications in Economics, Finance, and Sociophysics},  S\u{a}voiu, Gh., (Ed.) (pp. 149-168). Oxford: Academic Press.
  
 \bibitem{sousalmalarzgalam05}  Sousa, A.O., Malarz, K., Galam, S. (2005) Reshuffling spins with short range interactions: When sociophysics produces physical results. \emph{International Journal of Modern Physics C}, 16, 1507-1517.

  \bibitem{bischi2010global} Bischi, G.-I. \& Merlone, U. (2010). Global dynamics in adaptive models of collective choice with social influence, in  \emph{Mathematical modelling of collective behavior in socio-economic and life sciences},
Naldi, G., (Ed.)(pp. 223-244). Berlin:Springer.

\bibitem{daFontouraCosta} da Fontoura Costa, L. Oliveira Jr., O.N., Travieso, G. Rodrigues, F.A., Ribeiro Villas Boas, P., Antiqueira, L., Palhares Viana, M., Correa Rocha, L.E. (2011). Analyzing and modeling real-world phenomena with complex networks: a survey of applications. \emph{Advances in Physics} 60, 3.

\bibitem{Vega_redondo_book} Vega-Redondo, F. (2007) \emph{Complex Social Networks}. Cambridge:Cambridge University Press.

\bibitem{Bernasconi} Bernasconi, M. \& Galizzi, M. (2010). Network formation in repeated interactions: experimental evidence on dynamic behaviour. \emph{Mind Society }, 9, 193-228. 

\bibitem{BalaGoyal} Bala, V., Goyal, S. (2000). A Noncooperative Model of Network Formation. \emph{Econometrica}, 68, 1181–1229.

\bibitem{Toivonen} Toivonen, R., Onnela, J.-P., Saram\"aki, J., Hyv\"onen, J., Kaski, K. (2006). A Model for Social Networks. \emph{Physica A}, 371, 851–860.

\bibitem{BogunaPastor} Bogu\~n\'a, M., Pastor-Satorras, R., D\'iaz-Guilera, A., Arenas, A. (2004) Models of social networks based on social distance attachment. \emph{Physical Review E}, 70, 056122.

\bibitem{KumarNovak}  Kumar, R., Novak, J., Tomkins, A. (2010). Structure and Evolution of Online Social Networks. \emph{Link Mining: Models, Algorithms, and Applications}, 337-357

\bibitem{Pallanature} Palla, G., Barab\'asi, A.-L., Vicsek T. (2007).  Quantifying social group evolution. \emph{Nature}, 446, 664-667.

\bibitem{Kirman2} Copic, J., Jackson, M.O., Kirman, A. (2009). Identifying Community Structures from Network Data via Maximum Likelihood Methods, \emph{The B.E. Journal of Theoretical Economics} 9, 30.

 \bibitem{Kirman1} Kirman, A., Oddou, C., Weber, S. (1986). Stochastic communication and coalition formation. \emph{Econometrica}, 54, 129-138.
 
 \bibitem{Koulouris} Koulouris, A., Katerelos, I., Tsekeris, T. (2013).  Multi-Equilibria Regulation Agent-Based Model of Opinion Dynamics in Social Networks. \emph{Interdisciplinary Description of Complex Systems}, 11, 51-70.  

\bibitem{stauffer}
Stauffer, D. (2012). A Biased Review of Sociophysics. \emph{Journal of Statistical Physics}, 151, 9-20.

\bibitem{kirman} Kirman, A. (1992) Whom of What Does the Representative Individual Represent? \emph{Journal of Economic Perspectives}, 6, 117-136.

 \bibitem{3DalFornoMerlone08}  Dal Forno, A., Merlone, U. (2008). Network dynamics when selecting work team member, in Naimzada, A. K., Stefani, S., Torriero, A. (Eds.), \emph{Networks, Topology and Dynamics Theory and Applications to Economics and Social Systems, Series: Lecture Notes in Economics and Mathematical Systems}, 613, 229-240. 
  
 \bibitem{4DalFornoMerlone08ECO11} Dal Forno, A., Merlone, U. (2009). Social entrepreneurship effects on the emergence of cooperation in networks. \emph{Emergence: Complexity and Organization}, 11, 48-58.

 
\bibitem{1JSM} Lambiotte, R., Ausloos, M. (2007). Coexistence of opposite opinions in a network with communities. \emph{Journal of Statistical Mechanics},  P08026.
% $(http://www.arxiv.org/pdf/physics/0703266, 799kb )$

 \bibitem{RLMAJH}   Lambiotte, R., Ausloos, M., Holyst, J.A. (2007). Majority model on a network with communities, \emph{Physical Review E}, 75, 030101.
%$(http://www.arxiv.org/pdf/physics/0612146, 341kb)  $

 \bibitem{RLMA494LCNS3993}  Lambiotte, R. and Ausloos, M. (2006). Collaborative tagging as a tripartite network. \emph{Lecture Notes in Computer Science}, 3993 LNCS  III, 1114-1117.
%$ (http://arxiv.org/abs/cs.DS/0512090,  624kb)$

\bibitem{garcia} Garcia Cant\'u Ross, A.,  Ausloos, M. (2009). Organizational and dynamical aspects of a small network with two distinct communities: Neocreationists vs. Evolution Defenders. \emph{Scientometrics}, 80, 457-472.   %arXiv:0805.2912  

\bibitem{physaGR}  Rotundo, G., Ausloos, M. (2010). Organization of networks with tagged nodes and biased links: a priori distinct communities. The case of Intelligent Design Proponents and Darwinian Evolution Defenders. \emph{Physica A}, 389, 5479-5494.


\bibitem{Beckman} Beckman, M. (1952) A continuous model of transportation. \emph{Econometrica}, 20, 643-660.

\bibitem{Raa}  Ten Raa, T. (1986) The initial value problem for the trade cycle in euclidean space. \emph{Regional Science and Urban Economics}, 16, 527-546.

\bibitem{Puu} Puu, T. (1982). Outline of a trade cycle model in continuous space and time. \emph{Geographical Analysis}, 14, 1-9.

\bibitem{Meadows}  Meadows, D. L. (1970). \emph{Dynamics of Commodity Production Cycles}. Cambridge (MA):Wright-Allen Press.

\bibitem{Goodwin} Goodwin, B. C. (1965). Oscillatory behaviour in enzymatic control processes. \emph{Adv. in Enzyme Regulation}, 3, 425–438.

\bibitem{Murray}  Murray, J. D. (2002)\emph{ Mathematical Biology I. An Introduction 3$^{\small\mbox{rd}}$ edition}. Berlin:Springer. 

\bibitem{alison}  Heppenstall, A. J., Crooks, A. T., See,L. M., Batty M.(eds) (2012) \emph{Agent-Based Models of Geographical Systems}. Berlin:Springer

 \end{thebibliography}
\end{document}